\documentclass[twocolumn,showpacs,preprintnumbers,amsmath,amssymb]{revtex4}

\usepackage{graphicx}
\usepackage{dcolumn}
\usepackage{bm}

\newcommand{\disav}[1]{\left\langle #1\right\rangle}
\newcommand{\dis}[0]{\text{dis}}
\newcommand{\e}[0]{\text{e}}
\newcommand{\cd}[0]{\text{d}}
\newcommand{\cD}[0]{\mathcal{D}}
\newcommand{\cS}[0]{\mathcal{S}}
\newcommand{\cZ}[0]{\mathcal{Z}}
\newcommand{\ii}[0]{\text{i}}
\newcommand{\diszeta}[0]{\disav{\zeta}}
\newcommand{\erf}[0]{\text{erf}}
\newcommand{\larkin}[0]{L_{\text{p}}}

\newcommand{\deph}[0]{h_{\text{p}}}
\newcommand{\downto}[0]{\searrow}

\newcommand{\order}[1]{{\cal O}( #1)}

\begin{document}


\title{Perturbation theory
       for ac-driven interfaces in random media}

\author{Friedmar Sch\"utze}%
 \email{schuetze@thp.uni-koeln.de}
\affiliation{%
Institut f\"ur Theoretische Physik, Universit\"at zu K\"oln,
Z\"ulpicher Stra\ss e 77, 50937 K\"oln, Germany
}%

\date{\today}

\begin{abstract}
We study $D$-dimensional elastic manifolds driven by ac-forces in
a disordered environment using
a perturbation expansion in the disorder strength and
the mean-field approximation.
We find, that for $D\le 4$ perturbation theory produces non-regular 
terms that grow unboundedly in time.
The origin of these non-regular terms is explained. By using
a graphical representation we
argue that the perturbation expansion is regular to all orders 
for $D>4$. Moreover, for the 
corresponding mean-field problem we prove that ill-behaved diagrams
can be resummed in a way, that their unbounded parts mutually
cancel. 
Our analytical results are supported by numerical investigations.
Furthermore, we conjecture the scaling of the Fourier coefficients
of the mean velocity with the amplitude of the driving force $h$.
\end{abstract}

\maketitle

\section{\label{sec:int}%
Introduction%
}
The theoretical analysis of pinning phenomena of 
elastic objects in random potentials is an
important physical problem with a great impact on many fields of
research \cite{Kardar:PhRep98,Fisher:PhRep98}.
Elastic objects in disordered media subject to a constant
driving force are meanwhile well understood \cite{NB:AdvPh04}
at zero temperature and the influence of finite temperature has
been studied as well \cite{CGD:PRB00}.

In recent years, also the problem of ac-driven elastic
interfaces in disordered systems has
gained experimental 
interest \cite{Kleemann:PRL07,Kleemann:ARMR07,Jezewski:PRB08}.
In experiments, considerable attention is devoted to the behaviour
of the ac-susceptibility of ferroelectric thin films or ultrathin
ferromagnetic multilayers, which is believed to be related to the
motion of domain walls in these systems. So,
an interest to understand ac-driven domain wall motion in
disordered systems emerged. First results towards an explanation
of the hysteretic behaviour of the magnetisation 
\cite{LykNP:PRB99,NPV:PRL01} have been achieved and
the observed frequency regimes for the response of ferroic domain
wall motion are understood phenomenologically
\cite{FedMS:PRB04}.
An important contribution comes from the velocity 
hysteresis of moving domain walls \cite{GNP:PRL03}, but the
characterising parameters have not yet been worked out.
Generally, with regard to an analytic description of the
domain wall motion characteristics, results are scarce.
Yet, mainly numerical results have been obtained to
describe qualitative features of such systems, 
like the hysteresis or double-hysteresis of the velocity
\cite{GNP:PRL03,Glatz:phd,PGK:PRB04}.

In many cases, for a first quantitative analysis of 
complicated non-linear problems one uses a
perturbation expansion. 
However, perturbative approaches are
sometimes
hampered by mathematical subtleties, like non-analyticities
or singular perturbation theory (cf. \cite{BenderOrszag}),
or by physical obstacles such as non-perturbative excitations or strong coupling.
The difficulty with perturbation theory as a tool for the
analysis of pinned elastic objects has its own interesting history.

Until the beginning of the eighties, the lower critical dimension $d_l$
of the random field Ising model has been the subject of a long-lasting
debate. Dimensional reduction 
predicted that the lower critical dimension
equals $d_l=3$, whereas domain-wall arguments \cite{ImryMa:PRL75} lead to the conclusion, 
that $d_l=2$.
Eventually, in 1984 a final decision could be made and
dimensional reduction was proven to fail \cite{Imbrie:PRL84,Imbrie:CMP85}. 
The reason for the failure has been 
found later \cite{Fisher:PRL86, Villain:JPA88} to be connected with the existence of many 
metastable states for the domain walls separating different regimes in a multidomain
configuration. The plethora of metastable states arises from the 
dominance of the disorder over the domain wall
elasticity on length scales above the so-called Larkin length $\larkin$
\cite{Larkin:SovJETP70}
for sample dimensions $d<5$. A perturbative iteration to find the
energy minimum will not necessarily yield the correct extremal state
\cite{VillainSemera:JPL83,Engel:JPL85}.
Put in more mathematical terms, 
the formal perturbative treatment of the
domain walls assumes an analytic
disorder correlator. However, a functional
renormalisation group (FRG) treatment shows \cite{Fisher:PRL86}, that any initially 
analytic disorder
correlator develops a cusp-singularity at a finite length scale, which is
the Larkin length $\larkin$.

This insight has important consequences for the problem of an interface in a
disordered environment exposed to a constant driving force $h$.
At zero temperature and for small external force $h$, the interface
adjusts its configuration to balance the driving and the disorder, but
remains pinned and does not move on large time scales. If $h$ is tuned to exceed a critical
threshold $\deph$, then after transience has relaxed, the interface slides 
with a mean velocity $v$ that behaves
as $v\sim(h-\deph)^\beta$ for $h\downto\deph$. 
The system undergoes a non-equilibrium
phase transition, the so-called depinning transition, 
for which $v$ plays the role of an order parameter.
The critical behaviour close to the depinning transition has been investigated in
a number of works
\cite{NSTL:JP2F,NarayanFisher:PRB92,NarayanFisher:PRB93,
Ertas:PRE94,NSTL:APL,CDW:PRL01,CDW:PRB02,CDW:PRE04}.
Furthermore, the influence of finite temperature on this transition has
been considered in several articles \cite{CGD:PRB00,BKG:EPL08,Kolton:PRB09}.
Even the scaling behaviour
for small frequencies $\omega$ of an ac-drive on approaching the
critical point $(h,\omega)=(\deph,0)$ has been worked out \cite{GNP:PRL03}.

The depinning transition can, however, not be accounted for by perturbatively expanding the
disorder.
Therefore, perturbation theory is not applicable for small constant driving forces.

To achieve deeper quantitative insight into the ac-dynamics
of interfaces, 
perturbation theory seems to be the only feasible
method. 
Clearly, in the vicinity of the critical point $(h,\omega)=(\deph,0)$
perturbation
theory is expected to yield erroneous results, if not properly combined with 
an FRG treatment. Yet, FRG equations for the similar problem
of a constant driving force, have been obtained by the construction
of a perturbative series and a subsequent $\epsilon$-expansion.
So, understanding the perturbation theory is the first necessary 
starting point.
As has been said, pure perturbative calculations are restricted
to parameter regimes far away from the neighbourhood of the depinning transition.
Thus, for sufficiently large
frequencies and driving field amplitudes, from the physical
point of view there appears to be no contraindication against a
perturbative procedure.
However, as will become clear in this article,
for systems subject to periodic driving forces
perturbation theory in the disorder strength 
gives non-regular contributions to the velocity corrections for an internal
interface dimension $D\le 4$. With the attribute non-regular we refer
to expressions that grow unboundedly in time.
Such unbounded
contributions certainly do not reflect the true physical behaviour, 
but their origin deserves careful investigation.
Far away from the critical point, this behaviour of perturbation theory
cannot be related to the assumption of an analytic disorder correlator.
Quite the contrary, working with a cusped disorder correlator
brings additional difficulties due to the delta functions in its 
derivatives. 

In section \ref{sec:walls}, we introduce the equation of motion
for ac-driven elastic manifolds in disordered media, and
investigate its perturbative
expansion. 
Although our model is taylored to describe elastic manifolds, like
interfaces between two immiscible fluids or domain walls in ferroic systems,
we believe that our analysis also covers a wide range of
models for other interesting problems, 
e.g. charge density waves \cite{GruenerCDW:RMP88}
or flux lines in type-II superconductors \cite{Blatter:RMP94,GiaDou:PRL94,NatSchei:AdvPh00}.
After analysing the first non-vanishing order, we derive the diagrammatic
expansion to account for higher orders. This can be used to argue, that
perturbation theory works for $D>4$. The failure of perturbation theory
for $D\le 4$ is then analysed and explained. The well-known 
suitability of perturbation theory for estimates of the velocity of domain walls
driven by a constant force, far in the sliding regime, does not contradict
our statements for the ac-driving. We are going to take a look at this as well.
The technically involved and more mathematical treatments are taken out of
the main text and given in the appendices \ref{app:dg4proof} and 
\ref{app:hocalc}.

The failure of perturbation theory in all physically interesting cases $D\le 4$ 
underlines the importance of the mean-field approach,
which is the second central subject of this article.
In section \ref{sec:mf},
we explore the corresponding mean-field equation
of motion and its perturbative expansion. After some illustration
of the qualitative behaviour of the full solution, we prove, 
that the perturbative corrections are regular, i.e. they
remain bounded in all orders. The
bulk part of this inductive proof, the induction step, is outsourced to
appendix \ref{app:reg}.
Further, we show that for large enough driving field amplitudes,
sufficiently strong elastic coupling and high frequencies, 
the perturbative
results agree very well with the numerics for the full mean field equation
of motion. We conclude our considerations of the mean-field problem by
a numerical analysis of the lowest non-vanishing perturbative order, which
yields the decay law of the Fourier coefficients with the driving field 
strength. 

A table of frequently appearing symbols is provided in appendix \ref{app:symbols}.

\section{\label{sec:walls}%
The perturbative treatment of interfaces
in disordered media
}

\subsection{\label{sec:walls:model}%
The model%
}
Our analysis models interfaces and 
domain walls that are thin such that they can be described by
elastic $D$-dimensional manifolds, embedded
in a $D+1$ dimensional space. The manifold itself is
parametrised by a $D$-dimensional set $x$ of coordinates
and its position in space is given by $z(x,t)$.
We confine ourselves to the study of the zero-temperature
case, i.e. we do not take thermal noise into account.
Moreover, our model assumes small 
gradients and does not allow for overhangs. We expose the interface
to a periodic driving force
\begin{align}
\label{eq:acdrive}
h(t)=h\cdot\cos\omega t.
\end{align}
Then, the overdamped dynamics of elastic interfaces can be described by the
equation of motion
\begin{equation}
\label{eq:weom}
\gamma^{-1}\partial_tz(x,t)=\Gamma\nabla_x^2z+h(t)+
u\cdot g(x,z),
\end{equation}
which has already been introduced in earlier works 
\cite{Feigelman:JETP83,Bruinsma:PRL84,KL:PRB85}. 
In eq. (\ref{eq:weom}), $\Gamma$ and $\gamma$ are the stiffness and the inverse mobility
of the domain wall. For simplicity, we set $\gamma=1$ in the following.
The function $g(x,z)$ describes the quenched disorder, taken to be 
Gau\ss ian with the correlators given by
\begin{eqnarray}
\label{eq:wdisordera}
\disav{g(x,z)}&=&0\\
\label{eq:wdisorderc}
\disav{g(x,z)g(x',z')}&=&\delta^D(x-x')\Delta(z-z'),
\end{eqnarray}
where $\disav{\ldots}$ denotes the average over disorder.
The disorder correlator in $z$-direction is taken symmetric around 0
and decays exponentially on a length scale $\ell$. Further, we demand 
$\Delta(0)=1$, as the strength of the disorder shall be measured by $u$.
To be definite, we choose 
\begin{equation}
\label{eq:discorr}
\Delta(z-z')=\exp\left[-\left({z-z'\over\ell}\right)^2\right]
\end{equation}
in case we need a precise formula.
This choice corresponds to the case of an elastic manifold exposed to
random field disorder \cite{NSTL:APL}.
Throughout the whole paper we assume weak disorder. 
This means, that pinning forces are weak and the interface is pinned at the
fluctuations of the impurity concentration, and not at single pinning centres.
A more precise definition can be found e.g. in \cite{NSV:PRB90}.
For weak disorder, the random forces have to accumulate to overcome
the elasticity. On small length scales, elastic forces dominate and
the interface is essentially flat. By comparing the elastic and the disorder
term in (\ref{eq:weom}) one can estimate the length scale $\larkin$, called
the Larkin length, at which the two competing effects are of the same order.
The result is
\begin{equation}
\label{eq:larkin}
\larkin=\left[{\Gamma\ell\over u}\right]^{2\over 4-D}.
\end{equation}
Thus, the elastic term dominates on all length scales for $D>4$.

Finally, we specify the initial configuration for the equation of motion
(\ref{eq:weom}) to be a flat wall $z(x,t=0)\equiv0$.

\subsection{\label{sec:walls:perturbation}%
Perturbation theory - first order%
}
The equation of motion
(\ref{eq:weom}) can only be solved via an expansion in the
disorder strength $u$.
We are going to derive the perturbation expansion directly for
the equation of motion, since this appears simpler.
There is, however, another approach via functional integrals
which came in useful for the functional renormalisation group
calculations in the case of a constant driving force.
The reader who is more familiar with this technique may find
a brief treatment in appendix \ref{app:fiappr} which reveals the
connection to our methodology. 

The expansion is naturally performed around
the solution for the problem without disorder, i.e.
where $u=0$. In this case, we have a flat wall following the
driving field: $Z(t)=(h/\omega)\sin\omega t$.
Thus, we decompose $z(x,t)$ into the disorder-free solution and
a correction, i.e. $z(x,t)=Z(t)+\zeta(x,t)$.
The equation of motion
for the disorder correction $\zeta(x,t)$ is easily derived from
eq. (\ref{eq:weom})
\begin{equation}
\label{eq:pertweom}
(\partial_t-\Gamma\nabla_x^2)\zeta(x,t)=u\cdot g(x,Z+\zeta).
\end{equation}
The Green function for the differential 
operator on the left hand side is the well-known heat kernel
\begin{eqnarray}
\label{eq:pertprop}
(\partial_t-\Gamma\nabla_x^2)G(x,t)&=&\delta^D(x)\delta(t)\nonumber\\
G(x,t)&=&\Theta(t)\>
\int{\cd^Dk\over(2\pi)^D}\>\e^{\ii kx-\Gamma k^2t}.
\end{eqnarray}
The $k$-integral has to be cut off at some scale $\Lambda$, corresponding
to the inverse smallest length scale 
in the system.
To set up the perturbation series, we expand the correction in the disorder strength
\begin{equation}
\label{eq:wzetaexpand}
\zeta(x,t)=\sum\limits_{n=1}^\infty u^n\zeta_n(x,t)
\end{equation}
and the disorder force around the non-disordered solution
\begin{equation}
\label{eq:wdisexpand}
g(x,Z+\zeta)=\sum\limits_{k=0}^\infty \partial^{k}_2 g(x,Z){\zeta^k\over k!}.
\end{equation}
Thus, we obtain an equation for the first order correction:
\begin{equation}
\label{eq:w1storder}
\zeta_1(x,t)=\int\cd^Dx'\int\limits_0^\infty\cd t'\>G(x-x',t-t')g(x',Z(t')).
\end{equation}
Obviously, the disorder average vanishes. The disorder average for
the second order contribution is given by
\begin{align}
\disav{\zeta_2}(t)=&\int\limits_0^t\cd t_1\int\limits_0^{t_1}\cd t_2\>
\Delta'[Z(t_1)-Z(t_2)]\times\nonumber\\
&\int{\cd^D k\over(2\pi)^D}\>\e^{-\Gamma k^2(t_1-t_2)}.
\end{align}
The second order correction to the velocity follows straightforwardly
\begin{align}
\label{eq:v2ndo}
\disav{v_2}(t)=&\int\limits_0^{t}\cd t'\>
\Delta'[Z(t)-Z(t')]
\int{\cd^D k\over(2\pi)^D}\>\e^{-\Gamma k^2(t-t')}.
\end{align}
To get a first impression on how this expression behaves for large $t$,
we split off the Fourier-0-mode:
\begin{align}
\Delta'[Z(t)-Z(t')]&=F_0(\omega t)/\ell+p(t,t')\\
F_0(\omega t)&=
\sum\limits_{n=0}^\infty K_n\left({h\over\omega\ell}\right)\cdot
\sin (2n+1)\omega t.
\end{align}
Here, $p(t,t')$ is a well-behaved
oscillation around 0 in $t'$. Since $p(t,t')$ yields a bounded
contribution to $\disav{v_2}(t)$, 
we consider only $F_0(t)$ to find out, how
$\disav{v_2}$ increases asymptotically in time. The Fourier
coefficients $K_n$ can be determined analytically.
They are diminished when their argument increases or approaches zero and they
remain bounded. 
Integration over $t'$ yields
\begin{align}
\disav{v_2}(t)&\sim {F_0(\omega t)\over\ell}{S_D\over(2\pi)^D}\int\limits_0^\Lambda
{\cd k\over\Gamma}\>k^{D-3}\left[1-\e^{-k^2\Gamma t}\right]\\
&={t^{2-D\over 2}\over\ell}{F_0(\omega t)\over\Gamma^{D/2}}
\cdot a_D(t/\vartheta),
\end{align}
where
\begin{align}
\label{eq:2othetaint}
a_D(x)&={S_D\over(2\pi)^D}
\int\limits_0^{\sqrt{x}}\cd p\>p^{D-3}
\left[1-\e^{-p^2}\right].
\end{align}
For $x\to\infty$ the integral $a_D$
converges for $D<2$ and diverges logarithmically for $D=2$.
Thus, the asymptotic behaviour of the first perturbative correction in
time is given by
\begin{align}
\label{eq:v2oasympt1}
\disav{v_2}(t)&\sim c_D(t)\cdot\left\lbrace
\begin{matrix}
t^{2-D\over 2}&D<2\\
\log t&D=2\\
\text{const}&D>2
\end{matrix}
\right.
\end{align}
where $c_D(t)$ is some bounded function.
\begin{figure}
\includegraphics[width=\columnwidth]{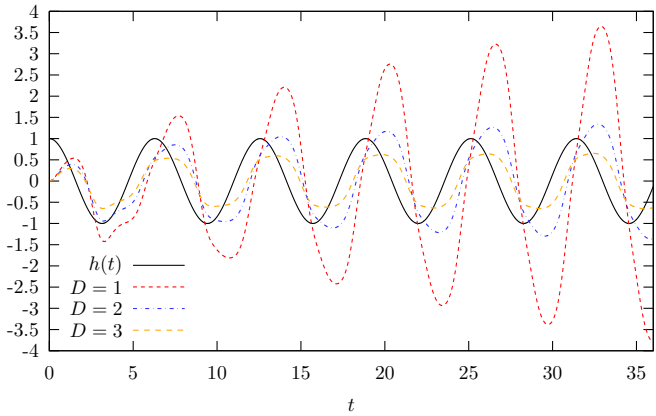}
\caption{\label{fig:wallsv} Plot of the first non-vanishing perturbative 
correction $\disav{v_2}(t)$ to the disorder average 
$\disav{v(x,t)}$ for different interface dimensions. For the plot we used
$h=1$ and the units are chosen such that $\omega=\ell=1$.}
\end{figure}%
Fig. \ref{fig:wallsv} shows the plots of $\disav{v_2}(t)$ for $D=1,2,3$.
Obviously, there is a problem of the perturbation expansion in low dimensions.
In the following, we are going to see that perturbation theory does not work
even for $D\le 4$.

\subsection{\label{sec:walls:highord}%
Higher order graphical expansion%
}

Higher orders of the perturbation expansion are best expressed
diagramatically. To deduce the diagrammatic rules, all one has to
do is plugging (\ref{eq:wzetaexpand}) into (\ref{eq:wdisexpand}),
rearranging the sum in powers of $u$ and inserting
this into (\ref{eq:pertweom}). The diagrammatic rules that emerge
are fairly simple: to express the perturbative correction of
order $n$, we draw all rooted trees with $n$ vertices and add a stem.
Up to the fourth order, this tree graph expansion is given by
\begin{eqnarray}
\label{eq:bentw1}
\zeta_1&=&\parbox{9mm}{\includegraphics{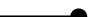}}
\\ &\phantom{=}&\nonumber \\
\zeta_2&=&\parbox{17mm}{\includegraphics{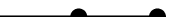}}
\\ &\phantom{=}&\nonumber \\
\zeta_3&=&
\parbox{17mm}{\includegraphics{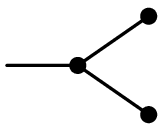}}\>+\>
\parbox{25mm}{\includegraphics{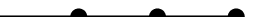}}
\\ &\phantom{=}&\nonumber \\
\zeta_4&=&
\parbox{17mm}{\includegraphics{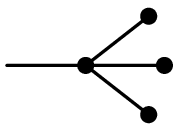}}\>+\>
\parbox{32mm}{\includegraphics{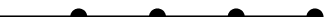}}\>+\nonumber
\\ &\phantom{=}& \nonumber \\
\label{eq:bentw4}
&&\parbox{25mm}{\includegraphics{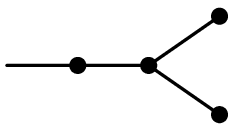}}\>+
2\cdot\>\parbox{25mm}{\includegraphics{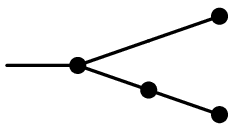}}
\end{eqnarray}
Every vertex represents a disorder insertion $g^{(m)}(Z(t))/m!$, where
$m$ counts the number of outgoing branches (away from the root).
Every line corresponds to an integral operator, the kernel being
the propagator $G(x,t)$. To get the graphical expansion for the
velocities, just remove the first line. The disorder average can be
carried out using Wick's theorem, since our disorder is assumed to be
Gau\ss ian. An immediate consequence is, that after averaging over disorder,
only graphs with an even number of vertices survive. In the following, we are going
to consider the perturbation expansion for the disorder averaged velocity $v(t)$.
The pairing for the disorder average shall be denoted by a dashed line. An example
graph from the $4^\text{th}$-order is
\begin{align}
\parbox{25mm}{\includegraphics{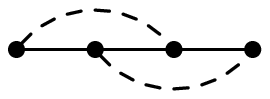}}.
\end{align}

Using this graphical expansion, in appendix \ref{app:dg4proof} 
we take a look at the general behaviour of the diagrammatic
contributions to all orders for $t\to\infty$ and argue that
all graphs remain bounded in case $D>4$.

\subsection{\label{sec:walls:dl4}%
The failure of perturbation theory for $D\le 4$
}
For $D\le 4$ the perturbation expansion is not as well-behaved as 
for $D>4$.
In this section, we show this and give an explanation why perturbation
theory is ill-behaved in low dimensions.

Though the disorder averaged
graphical structure can
become complicated, one especially simple graph
has the same structure in all orders: the one for which all vertices are 
connected directly to the root. 
In the equations (\ref{eq:bentw1}-\ref{eq:bentw4}),
we have drawn those graphs at the very first place. Let us call them 
bushes. The general bush graph contribution to the velocity correction
of order $2p$ thus corresponds to the following diagram
\begin{align}
\disav{B_{2p}}&=\>\parbox{13mm}{\includegraphics{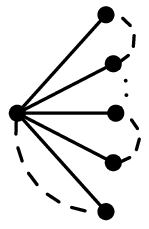}},
\end{align}
where the dotted line is a placeholder for other vertices that we have not
drawn.
Up to combinatorical factors, the general (disorder averaged) 
bush $B_{2p}$ that occurs in the $2p$-th order perturbative 
correction to the velocity $v$, reads
\begin{align}
\label{eq:buschallg}
\disav{B_{2p}}&(t)\propto\nonumber\\
&\bigg[\int\limits_0^t\cd t_1\cd t_2\>
\Delta[Z_1-Z_2]
\int{\cd^Dk\over (2\pi)^D}\>\e^{-\Gamma k^2(2t-t_1-t_2)}
\bigg]^{p-1}\nonumber\\
&\times
\int\limits_0^t\cd t'\>\Delta^{(2p-1)}[Z-Z']
\int{\cd^Dk\over (2\pi)^D}\>\e^{-\Gamma k^2(t-t')}\\
\label{eq:t1t2def}
&\equiv T_1^{p-1}\cdot T_2
\end{align}
To work out the asymptotic envelope of $\disav{B_{2p}}(t)$ for large $t$
we only need to take a look at the expression $T_1$.
Since $\Delta$ is an entirely positive function, and the $k$-integral
is certainly non-negative, we can replace $\Delta$ by its maximum
$\Delta(0)$ to get an upper bound, and if we 
replace $\Delta$ by the minimal value that is taken 
$\Delta[2h/(\omega\ell)]$, we get a lower bound for $T_1$. In both cases,
$\Delta[Z(t_1)-Z(t_2)]$ is replaced by a constant, so that the integration over
$t_1$ and $t_2$ can be done easily. Consequently, up to a constant, 
the asymptotic envelope of $T_1$ is given by
\begin{align}
\label{eq:t1absch}
{u^2\over\ell^2}T_1(t)\sim&
{u^2\over\ell^2}\int{\cd^Dk\over (2\pi)^D}{\left[1-\e^{-\Gamma k^2t}\right]^2\over
\Gamma^2k^4}\nonumber\\
=&\left[{t\over\tau}\right]^{4-D\over 2}A_D(t/\vartheta).
\end{align}
Here, the constants $\tau$ and $\vartheta$ are
time scales, given by
\begin{align}
\label{eq:tautscale}
\tau&=\Gamma^{D\over 4-D}\left[{\ell\over u}\right]^{4\over 4-D}=\left({\ell\over u}\right)
\larkin^{D\over 2}=\larkin^2/\Gamma\\
\label{eq:thetatscale}
\vartheta&=\Lambda^{-2}/\Gamma
\end{align}
and the function $A_D(x)$ has been introduced for notational convenience
\begin{align}
\label{eq:ADfunc}
A_D(x)={S_D\over (2\pi)^D}
\int\limits_0^{\sqrt{x}}\cd p\>p^{D-5}\left[1-\e^{-p^2}\right]^2.
\end{align}
The function $A_D(x)$ is increasing but bounded for $D<4$ and grows logarithmically as
$x\to\infty$ for $D=4$. Thus $T_1$ grows monotonically in $t$ for any $D\le 4$.

Actually, our statement about the asymptotics in eq. (\ref{eq:t1absch}) 
is more robust than our simple argument suggests, and in fact does not 
rely on the positivity of $\Delta$. A more detailed calculation,
presented in appendix \ref{app:hocalc} shows that
\begin{align}
{u^2\over\ell^2}T_1=&
\left({t\over\tau}\right)^{4-D\over 2}
\kappa_D\left({t\over\vartheta}\right)+
\nonumber\\
&\left({t\over\tau}\right)^{4-D\over 4}
{u\Lambda^{D\over 2}\over\omega\ell}
k_D\left({t\over\vartheta},\omega t\right)+
\nonumber\\
&{u^2\Lambda^D\over\omega^2\ell^2}
P_D\left({t\over\vartheta},\omega t\right)
\end{align}
and
\begin{align}
{u^2\ell^{2(p-1)}\over\omega\ell}T_2=&
{u\Lambda^{D\over 2}\over\omega\ell}
\left({t\over\tau}\right)^{4-D\over 4}
f_D\left({t\over\vartheta},\omega t\right)+
\nonumber\\
&{u^2\Lambda^D\over\omega^2\ell^2}
p_D\left({t\over\vartheta},\omega t\right)
\end{align}
where all of the functions $\kappa_D$, $k_D$, $P_D$, $f_D$ and $p_D$
are bounded in $t$ for $D<4$. 
The first term of $T_1$ shows, that for $D\le 4$
the whole expression grows for $t\to\infty$
without any bound. This is, because by definition (cf. equation (\ref{eq:kappad})),
$\kappa_D\propto A_D$, where $A_D$ is given by (\ref{eq:ADfunc}).
The other tree graphs that appear in the graphical representation of $\disav{v_{2p}}$
and that we have not analysed here, exhibit similar behaviour. 
Cancellations among diagrams do not occur.
To have a little more evidence, 
that $D=4$ really enters as an upper critical dimension,
we have perturbatively investigated the interface's width in appendix \ref{app:width}.
In summary, perturbation theory is ill behaved for $D\le 4$. The reason for this
shall be discussed in the following.

The bushes $\disav{B_{2p}}$ that we have considered so far are part of an expansion
of the disorder averaged velocity 
\begin{align}
v(t)=h\cdot\cos\omega t+v_\text{dis}(t)
\end{align}
in the disorder
strength $u$. The dimensionless ratios, in which $u$ occurs in that expansion
are
\begin{align}
\left({t\over\tau}\right)^{4-D\over 4}\quad
\text{and}\quad
{u\Lambda^{D\over 2}\over\omega\ell}.
\end{align}
Since in the stationary state the interface is expected to follow the driving
with frequency $\omega$, its Fourier representation has to take on the form
\begin{align}
v_\text{dis}(t)=
\omega\ell\sum\limits_n&e_n\left(\left({t\over\tau}\right)^{4-D\over 4},
{u\Lambda^{D\over 2}\over\omega\ell},{t\over\vartheta}\right)
\cos n\omega t+\\
&f_n\left(\left({t\over\tau}\right)^{4-D\over 4},
{u\Lambda^{D\over 2}\over\omega\ell},{t\over\vartheta}\right)
\sin n\omega t.
\end{align}
Because of the ratio $t/\tau$, an expansion in the disorder $u$ brings
powers of $t^{4-D\over 4}$ in every order, since $\tau$ depends on $u$ (cf. equation 
(\ref{eq:tautscale})). The remaining question is the meaning of $\tau$.
Since $\tau$ appears as a time scale for the time dependence of the Fourier
coefficients, which physically should approach a constant value in the
stationary state $t\to\infty$, the most natural interpretation is the transience.
Keeping in mind, that we start with a flat wall, we have to expect several
kinds of transience effects. As we have seen, there are only two time scales
in question: $\tau$ and $\vartheta$. As can be concluded from their definitions
(cf. equations (\ref{eq:tautscale}) and (\ref{eq:thetatscale})), they obey
the same structure: an intrinsic length scale to the power 2 divided by $\Gamma$.
The time scale $\tau$ involves the Larkin length $\larkin$ (cf. equation 
(\ref{eq:larkin})), which measures the competition between disorder and
elasticity: the interface is flat on length scales $L\lesssim\larkin$. 
This indicates that, up to some dimensionless prefactor, 
$\tau$ describes the time during which correlated interface segments of 
extension $\larkin$ adopt to their local disorder environment, i.e. the 
roughening time of
the interface. For $D>4$, the interface is flat on all length scales, thus 
there is neither roughening nor a Larkin length, hence $\tau$ is 
meaningless
and cannot occur. This agrees with our observation. For $D>4$, there is no
disorder-dependent time scale any more, which could bring powers of time
in the perturbative corrections, therefore they remain finite as $t\to\infty$.
However, also for $D>4$ the disorder leads to a typical
deviation of every point of the interface from the mean position. 
The built-up of this typical deviation towards its steady-state value 
happens on the time-scale $\vartheta$, in agreement with the observation
for mean-field theory (cf. section \ref{sec:mf:pert}).
Thus, both time scales can naturally
be interpreted as the life times of two different transience effects.

\subsection{\label{sec:walls:dc}%
Interfaces subject to a constant driving force%
}
In the introduction, we already pointed out that perturbation theory in
connection with interfaces driven by a constant force cannot
properly account for the
existence of the depinning transition and therefore gives misleading 
results for $D\le 4$.
But far above the depinning threshold, i.e. for $h\gg\deph$, the interface 
slides and its
velocity can be estimated perturbatively. The dynamical correlation length
$L_v=\Gamma\ell/v$ \cite{NSTL:APL} is then small compared to the Larkin length 
$\larkin$ and thus 
working with an analytic disorder correlator and expanding the
disorder in its moments works. Of course, also for constant driving forces
one has to start with a definite initial condition, which is usually a
flat wall. Thus, there will be transience effects for dc-driven interfaces 
as well \cite{SchehrDou:EPL05,KoSchehrDou:PRL09}.
In this section we take a short glance
to understand the difference between ac and dc-driving. Our special interest
is devoted to the time scales that determine the duration of transience
effects. 

The equation of motion for the elastic interface experiencing a constant 
driving force
\begin{equation}
\label{eq:dceom}
\partial_tz(x,t)=\Gamma\nabla_x^2z+h+u\cdot g(x,z),
\end{equation}
has the disorder-free solution ($u=0$) $Z(t)=ht$. The perturbation expansion is essentially
the same as in section \ref{sec:walls:perturbation}, just the non-disorded solution
around which we expand is different.

Actually, there is a problem with the decomposition $z=Z+\zeta$ here, since the
sliding velocity $v$ is different from $h$, hence $\zeta\sim(v-h)t$ is
not a small quantity (compared to $\ell$) for large $t$ 
and the Taylor expansion (\ref{eq:wdisexpand}) of the disorder is questionable.
Since here we shall not be interested in large times $t>\ell/(h-v)$
but only want to
determine the time scale of the transience (occuring at small $t\ll\ell/(h-v)$), 
this problem is safely ignored.

The first non-vanishing correction to the
velocity is found to be (cf. eq. (\ref{eq:v2ndo}))
\begin{align}
\disav{v_2}(t)=&\int\limits_0^{t}\cd t'\>
\Delta'[Z(t)-Z(t')]
\int{\cd^D k\over(2\pi)^D}\>\e^{-\Gamma k^2(t-t')}\\
=&{1\over h}\int{\cd^D k\over(2\pi)^D}
\left[\e^{-k^2\Gamma t-{h^2t^2\over\ell^2}}
-\varphi\left({\Gamma\ell k^2\over 2h},{ht\over\ell}\right)
\right]\\
=&{\Lambda^D\over (t/\vartheta)^{D\over 2}h}{S_D\over (2\pi)^D}
\int\limits_0^{\sqrt{t/\vartheta}}\cd p\>p^{D-1}\times\nonumber\\
&\left[\e^{-p^2-
{{h^2t^2\over\ell^2}}}-
\varphi\left({p\over 2}{{\ell\over ht}},{{ht\over\ell}}\right)\right]
\end{align}
where we have introduced the function
\begin{align}
\varphi(a,b)=1-\sqrt\pi\,a\cdot\e^{a^2}\cdot\big[\erf(a+b)-\erf(a)\big]
\end{align}
for convenience.
The time-scales on which transience effects disappear are obviously given by
\begin{align}
\tau_\text{dc}={\ell\over h}={L_h^2\over\Gamma}\quad\text{and}\quad
\vartheta={\Lambda^{-2}\over\Gamma}
\end{align}
and are manifestly disorder-independent. 
$L_h=\Gamma\ell/h$ denotes the correlation length of an interface moving with a velocity
$h$. 
Of course, asymptotically the interface moves with a velocity
$v<h$, but at the very beginning, when we start off with a flat wall, its
velocity is indeed given by $h$ (cf. (\ref{eq:dceom})). 

It is not surprising, that
the time scale of the initial roughening is different for dc and ac
driving. Either problem involves completely different physical
processes to be responsible for transience. In the case of an
ac-driving, the system undergoes a process of adaption of its
configuration to the \emph{local} disorder, such that a 
stable stationary oscillation is possible, and during which higher
Fourier modes build up.
For dc-driving, the system starts to move with a velocity of
$h$, which then rapidly decreases, and roughens
since segments of the interface are pinned and remain at rest until 
they are pulled forward by the neighbouring segments through the
elastic coupling. The time for this process mainly depends on the
velocity of the interface, not on the strength of the disorder.

\section{\label{sec:mf}%
Mean field theory%
}

To extend the study of ac-driven interfaces beyond numerics, 
perturbation theory seems unavoidable. 
Since, as we have seen in the preceeding section, perturbation
theory only works for $D>4$, we turn here to an interesting limiting
case, formally corresponding to $D=\infty$: the mean-field equation of motion.

Mean field calculations have been performed before for the problem of
interfaces and charge density waves subject 
to a constant driving in a number of articles
\cite{Fisher:PRL83,Fisher:PRB85,Leschorn:JPA92,NarayanFisher:PRB92,Lyuksyutov:JPC95}.
Special emphasis has been put on the depinning transition and
its critical properties.

The perturbation expansion for dc-driven interfaces has been 
investigated by Koplik and Levine \cite{KL:PRB85}, who also emphasised on the
mean-field problem. They already mentioned the same problematic graphs in their
expansion that we will encounter below, but they did not provide a proof for the fact, 
that the unbounded terms cancel to all orders, independent of the driving.

\subsection{\label{sec:mf:model}%
The mean-field equation of motion%
}

The mean field equation corresponding to our original equation of
motion (\ref{eq:weom})
is obtained via the replacement of the elastic term
by a uniform long-range coupling (cf. e.g. \cite{Fisher:PRB85}). To do this,
we have to formulate the model (\ref{eq:weom}) on a lattice in
$x$-direction, i.e. the coordinates that parameterise the interface itself 
are discretised.
The lattice Laplacian reads
\begin{align}
\nabla_x^2z(x_i)&=
\sum\limits_{d=1}^D{z(x_i+ae_d)+z(x_i-ae_d)-2z(x_i)\over a^2}
\nonumber\\
&=\sum\limits_{d=1}^D\sum\limits_{j_d=1}^NJ_{ij_d}\big[z(x_{j_d})-z(x_i)\big],\\
J_{ij_d}&={1\over a^2}\big[\delta_{j_d+1,i}+\delta_{j_d-1,i}\big],
\end{align}
where $a$ denotes the lattice constant.
To get the mean field theory, $J_{ij}$ has to be replaced by a uniform coupling but
such that the sum over all couplings $\sum_jJ_{ij}$ remains the same. Hence, we choose
\begin{align}
J_{ij}^{\rm MF}={1\over a^2N}.
\end{align}
Now, the disorder has to be discretised as well, which is achieved if we
replace the delta function in the correlator (\ref{eq:wdisorderc}) by
$\delta^D(x_i-x_j)\to \delta_{ij}a^{-D/2}$ (cf. \cite{Bruinsma:PRL84}).
The resulting equation of motion should be independent of $x$, just the 
lattice constant $a$ and the
dimension enter because the disorder scales with a factor $a^{-D/2}$. 
Finally, for the mean-field equation of motion, we obtain
\begin{align}
\label{eq:mfgeom}
\partial_tz
&=c\cdot\left[\disav{z}-z\right]+h(t)+
\eta\cdot g(z),
\end{align}
where $c=\Gamma/a^2$ and $\eta=u/a^{D/2}$.
The disorder remains Gau\ss ian with
\begin{eqnarray}
\label{eq:disorder}
\disav{g(z)}&=&0\\
\disav{g(z)g(z')}&=&\Delta(z-z').
\end{eqnarray}
The function $\Delta(z-z')$ is as before, so we shall 
choose again (\ref{eq:discorr})
whenever we need an explicit expression for calculations.

The physical picture of the
mean field equation of motion is a system of distinct particles,
moving in certain realisations of the disorder. All of them are
harmonically coupled to
their common mean, i.e. the elastic coupling between neighbouring
wall segments $\Gamma\nabla_x^2z$ is now replaced by a uniform coupling
$c\cdot\left[\disav{z}-z\right]$ to 
the disorder averaged position $\disav{z}$, which in turn is determined self-consistently 
by the single realisations.

Apart from the correlation length $\ell$ of the disorder, there is another
important length scale in the system. In the absence of any
driving force (i.e. $h=0$), we can easily determine the
mean deviation of the coordinate
$z$ of a special realisation from the disorder averaged position 
$\disav{z}$. For $h=0$ we expect $\dot z=0$, at least in the steady state
and (\ref{eq:mfgeom}) straightforwardly leads to 
\begin{align}
\label{eq:meandev}
\disav{(\disav{z}-z)^2}\simeq {\eta^2\over c^2}.
\end{align}
So, $\eta/c$ measures the order of the average distance from the common mean.

In what follows, the disorder averaged velocity $v=\disav{\dot z}$ will
be denoted by the symbol $v$.

\subsection{\label{sec:mf:numerics}%
Qualitative behaviour and numerical results%
}
To get an idea about how the system, corresponding to the equation
of motion with an ac-driving (cf. (\ref{eq:mfgeom}))
\begin{align}
\label{eq:mfacfeld}
h(t)=h\cdot\cos\omega t
\end{align}
behaves, we implemented a numerical approach. The disorder is modelled
by concatenated straight lines, the values of the junction points are 
chosen randomly from a bounded interval. The correlator has been checked to
be perfectly in agreement with (\ref{eq:discorr}).

\begin{figure}
\includegraphics[width=0.9\columnwidth]{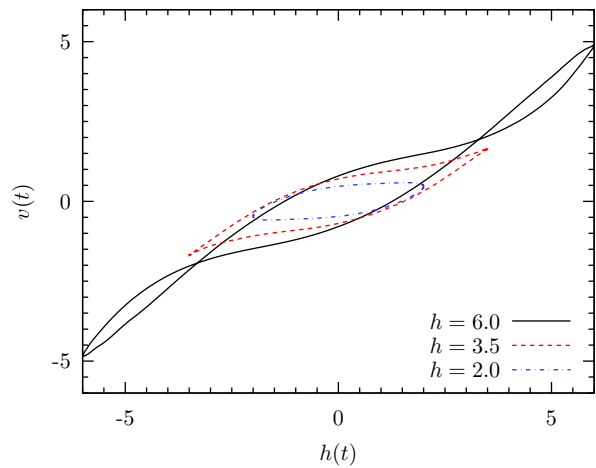}
\caption{\label{fig:qbild} Numerical solution of (\ref{eq:mfgeom})
for different driving field strengths and $c=1.0$, $\eta=2.5$.
For the simulation, $t$ and $z$ are measured in units such that
$\omega=\ell=1$.}
\end{figure}%
Before discussing the numerical trajectories, we note a first property
of the equation of motion (\ref{eq:mfgeom}). It contains a symmetry of the
(disorder averaged) system, namely that all disorder averaged
quantities are invariant under the 
transformation $h\to -h$ 
and $z\to -z$, which implies $v\to -v$. We have hereby fixed the initial
condition to be $z(0)=0$ for all realisations. If one chooses another
initial condition, its sign has to be inverted as well, of course.
In the steady state, i.e. for $t\gg c^{-1}$ (as $c^{-1}$ is the 
time-scale on which transience effects are diminished, see below), the
trajectory must therefore obey the symmetry $h\to -h$, $v\to-v$.
This symmetry is obviously reflected in the numerical solutions 
(see fig. \ref{fig:qbild}).

An interesting consequence of this symmetry is, 
that the even Fourier coefficients of
the solution $v(t)$ (which is periodic with period $2\pi/\omega$)
vanish. Once the steady state is reached, the symmetry requires
$v(t)=-v(t+\pi/\omega)$. For the even Fourier modes this means
\begin{align}
c_{2N}&=\int\limits_0^{2\pi\over\omega}\cd t\>v(t)\e^{\ii2N\omega t}
\nonumber\\
&=\int\limits_0^{\pi\over\omega}\cd t\>v(t)\e^{\ii2N\omega t}+
\int\limits_0^{\pi\over\omega}\cd t\>v\left(t+{\pi\over\omega}\right)
\e^{\ii2N\omega t}
=0.
\end{align}

The typical picture of a $v$-$h$-plot is that of a single hysteresis
for $h\ll\eta$ and  a double hysteresis for $h\gg\eta$. In an intermediate
range, we find a single hysteresis with a cusped endpoint. The qualitative
shape of the solution trajectories, examples of which are shown in
fig. \ref{fig:qbild}, agrees with numerical 
results \cite{GNP:PRL03,Glatz:phd}, that
have been obtained as solutions for (\ref{eq:weom}) in the case of finite
interfaces with periodic boundary conditions.
\begin{figure}
\includegraphics[width=\columnwidth]{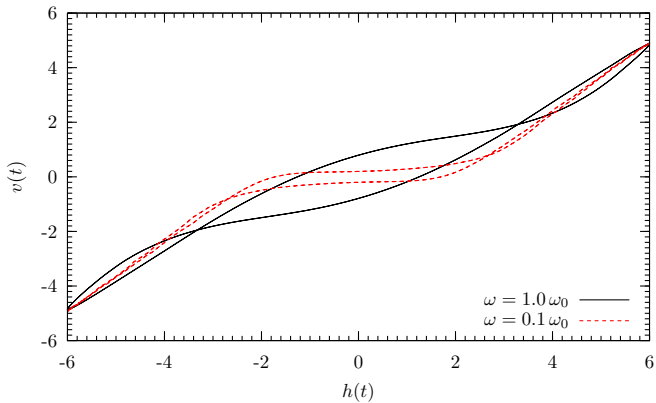}
\caption{\label{fig:wto0} Numerical solution of equation (\ref{eq:mfgeom})
for $h=6.0$, $c=1.0$ and $\eta=2.5$ for different frequencies, 
$t$ and $z$ being measured in units such that
$\omega_0=\ell=1$.}
\end{figure}%
Moreover, as the frequency is sent to zero $\omega\to 0$, the hysteretic
trajectory approaches the depinning curve for an adiabatic
change of the driving field. This is shown in fig. \ref{fig:wto0}.

In the following, we want to give a qualitative discussion of the 
hystereses in the case of small elasticity $c$, more precisely our
discussion assumes $\eta/c\gg\ell$.

\emph{Weak fields $h\ll\eta$:}\quad
For each disorder realisation, the configuration is
allowed to deviate from the mean by an order of $\eta/c$ (cf. (\ref{eq:meandev})), 
which by our assumption is large compared to the disorder correlation length $\ell$. 
So, except for
rare events, in the case of weak driving fields, the typical system in a certain
disorder configuration remains in a potential well of the disorder.
Starting at some large time $t_0$ (in the steady state) for which
$h(t_0)=0$,
we expect a certain realisation to be located close to a zero
point $z_0$ of a falling edge of the disorder force $g(z_0)=0$, 
since this corresponds to a stable
configuration. As the field grows, the system starts to move in the
direction of growing $z$, where the disorder force competes with
the driving. Because in the vicinity of the potential minimum,
the disorder force $g(z)$ behaves approximately
linear in $z$, the acceleration is approximately zero and
the velocity almost constant. This changes when the driving is about to 
reaching its maximum. The slower the growth of the driving, the smaller the velocity.
At the maximum, the velocity equals zero, as the driving and the restitutional 
disorder force compensate. For decreasing $h(t)$, the restitution force
wins and pushes the system back in the direction of the potential minimum. 
Hence, the velocity $v$
turns negative short-time after the field has reached its maximum and is still positive.
Once the stable position $z_0$ is reached again, 
the same starts in the negative direction.

Certainly, the restitutional disorder force need not continuously grow with 
$z$, but may exhibit bumps or similar noisy structure, but those details average 
out when taking the mean over all disorder configurations.

\begin{figure}
\includegraphics[width=\columnwidth]{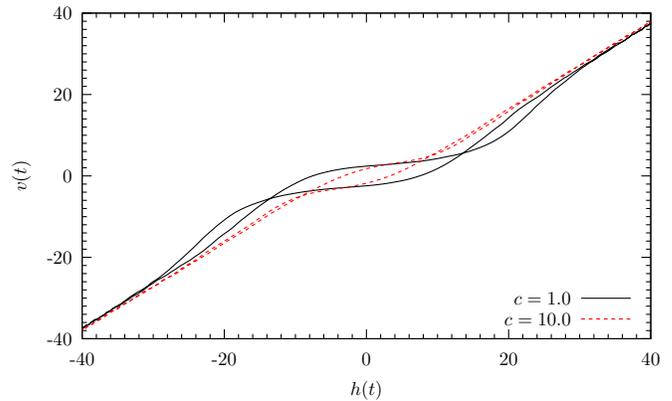}
\caption{\label{fig:ceinfl} Numerical solution of equation (\ref{eq:mfgeom})
for different elastic constants and $h=40.0$, $\eta=10.0$.
The units of $t$ and $z$ are chosen such that
$\omega=\ell=1$.}
\end{figure}

\emph{Strong fields $h\gg\eta$:}\quad
In the case of strong driving amplitudes, we encounter the situation of a
double hysteresis. Again, starting at a time $t_0$ with $h(t_0)=0$ 
for $t_0\gg c^{-1}$, we may assume the system to be
located close to a zero $g(z_0)=0$ of a falling edge of the
disorder force field. As $h(t)$ grows, we first have the same situation as in
the case of weak driving: The disorder acts restitutionally and thus keeps the 
velocity small and leads to a small slope $\cd v/\cd h$. Once the field
is of the order $\eta$, the system is no longer
locked into a potential well, but a cross-over to sliding behaviour sets in. 
On further increasing $h$, the
system finally arrives at a slope $s=\cd v/\cd h$, which depends mainly
on $\eta$ and $h$. The larger $h$, the closer $s$ approaches the value of 1.
After the field reaches
its maximum, the velocity decreases with the field, the slope being 
smaller than $s$, if $h$ is not too large. 
This is explained as follows. The mean deviation
$\disav{(\disav{z}-z)^2}$ that we have estimated to be of the order $\eta^2/c^2$
for $h=0$ (cf. eq. (\ref{eq:meandev})),
under driving also depends on the difference between $v(t)$ and $h(t)$
(as can be seen from (\ref{eq:mfgeom})).
At the beginning of the crossover to sliding, when $h\approx\eta$, this difference
is large and therefore also the mean deviation is large (especially for small $c$). 
When $h$ increases, the
particles in disorder realisations that are behind the mean are
strongly accelerated. On the other hand, those that are ahead of the mean position
are not so much decelerated, because for each disorder
realisation, being ahead of or behind the mean position rapidly changes and
a motion backwards (against the driving) is suppressed. Thus
the reduction of the mean deviation towards some asymptotic value gives a
stronger slope for rising fields during the crossover. When the field $h(t)$
reduces, the mean deviation shrinks, $h(t)$ approaching $v$, and $v$
drops less rapidly than it has risen because still the mean deviation prefers 
a diminution of the difference $v(t)-h(t)$.
This approximately linear reduction of the velocity remains, until
the field is weaker than the typical disorder force, when the system is
again trapped in a potential well. Since on rising edges of the disorder force,
driving and disorder point in the same direction, the system will rarely
sit there (it moves away very fast). The velocity becomes negative before
$h=0$, since the system slides down the falling edge ($\cd g/\cd z<0$) 
of the disorder force.
At $h=0$ everything starts again in the negative direction.
An example for fairly large
field amplitudes is shown in fig. \ref{fig:ceinfl}.

So far, our discussion has emphasised on small $c$. 
The effect of larger $c$ is to couple the 
configuration $z(t)$ of every realisation strongly to the mean $\disav{z(t)}$.
This wipes out the effect of disorder. 
Thus for larger $c$ the double hysteresis winds around a 
straight line, connecting the extremal velocities. This can be seen in
fig. \ref{fig:ceinfl}.

\subsection{\label{sec:mf:pert}%
Mean-field perturbation theory%
}

\subsubsection{\label{sec:mf:pert:diag}%
The diagrammatic expansion%
}

Since the mean-field equation of motion (\ref{eq:mfgeom}) cannot
be solved exactly, we attempt an expansion
in the disorder strength $\eta$. Therefore, as before, we decompose $z=Z+\zeta$, where 
$Z(t)=(h/\omega)\sin\omega t$ is
the solution of the non-disordered problem ($\eta=0$) around which we expand, and 
\begin{align}
\zeta=\sum\limits_{k=1}^\infty\zeta_k\eta^k\>,\quad
\diszeta=\sum\limits_{k=1}^\infty\diszeta_k\eta^k.
\end{align}
is the perturbative correction.
Still, we have the equations for $\zeta_k$ depending on $\diszeta_k$,
which is also unknown. This eventually leads us to a set of two coupled
equations
\begin{align}
\label{eq:mfpertsys1}
(\partial_t+c)\zeta&=c\diszeta+\eta\cdot g(Z+\zeta)\\
\label{eq:mfpertsys2}
\partial_t\diszeta&=\eta\cdot\disav{g(Z+\zeta)},
\end{align}
that we can solve iteratively for every order of the perturbation 
series, if we expand
\begin{align}
\label{eq:mfdisexpansion}
g(Z+\zeta)=\sum\limits_{n=0}^\infty{g^{(n)}(Z)\over n!}\zeta^n.
\end{align}
If one is interested to keep small orders, this expansion of the disorder 
can only work if $\zeta\ll\ell$, because $\ell$ is the typical scale on which $g(z)$
changes. 
We will come back to that point later in section \ref{sec:mf:pert:valpert}, when 
discussing the validity of perturbation theory.
For the moment, we just do it.

The propagator corresponding to the left hand side of Eq. (\ref{eq:mfpertsys1}) reads
\begin{align}
\label{eq:mfprop}
G(t)=\Theta(t)\cdot\e^{-ct}.
\end{align}
Using this propagator, we can formally write down the solution and express
it order by order in a power series in $\eta$.
Up to the second order, the solutions are
\begin{align}
\diszeta_1(t)&=0,\\
\zeta_1(t)&=\int\limits_0^t\cd t_1\>\e^{-c(t-t_1)}g(Z(t_1)),\\
\label{eq:diszeta2}
\diszeta_2(t)&=\int\limits_0^t\cd t_1\int\limits_0^{t_1}\cd t_2
\>\e^{-c(t_1-t_2)}\Delta'[Z(t_1)-Z(t_2)],\\
\label{eq:zeta2}
\zeta_2(t)&=\int\limits_0^t\cd t_1\e^{-c(t-t_1)}\left[c\diszeta_2(t_1)+g'(Z(t_1))
\cdot\zeta_1(t_1)\right].
\end{align}%
Since we assume Gau\ss ian disorder, the disorder averaged corrections
$\diszeta_n$ vanish for odd $n$.
We use a diagrammatic representation to depict the nested perturbation 
expansion. For the interesting quantities $\diszeta_k$, 
the first two non-vanishing orders are given by: 
\begin{widetext}
\begin{align}
\label{eq:diagexpand}
\diszeta_2=&\quad
\parbox{17mm}{\includegraphics{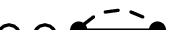}}
\nonumber\\&\phantom{=}\nonumber\\
\diszeta_4=&\quad 
3\cdot\parbox{17mm}{\includegraphics{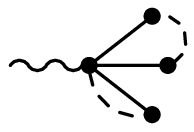}}\quad+\quad
\parbox{25mm}{\includegraphics{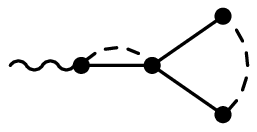}}\quad+\quad
2\cdot\parbox{25mm}{\includegraphics{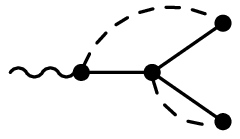}}\quad+
\nonumber\\&\phantom{=}\\&\quad
2\cdot\parbox{25mm}{\includegraphics{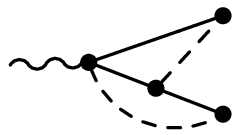}}\quad+\quad
2\cdot\parbox{25mm}{\includegraphics{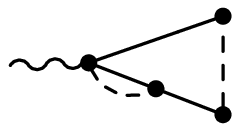}}\quad+\quad
2\cdot\parbox{25mm}{\includegraphics{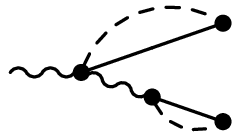}}\quad+
\nonumber\\&\phantom{=}\nonumber\\&\quad
\parbox{33mm}{\includegraphics{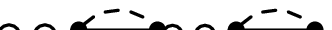}}\quad+\quad
\parbox{33mm}{\includegraphics{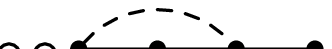}}\quad+\quad
\parbox{33mm}{\includegraphics{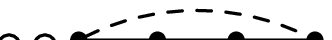}}
\nonumber
\end{align}
\end{widetext}
The diagrammatic rules are fairly similar to those in section \ref{sec:walls:highord}:
we draw all rooted trees with $k$
vertices, and add a stem. Each vertex
corresponds to a factor $g^{(m)}(Z(t))/m!$, where $m$ counts the number of
outgoing lines (away from the root). 
The line between two vertices represents a propagator $G(t)$.
Then Wick's theorem is applied to carry out the disorder average. Each two vertices, that
are grouped together for the average, will be connected by a dashed line.
Finally, there is one new feature, that we did not come along in section
\ref{sec:walls:highord}. Every straight line which, upon removing it, makes the whole
graph falling apart into two subgraphs, has to be replaced by a curly line. 
A curly line symbolises the propagator of (\ref{eq:mfpertsys2}), which is just a Heaviside
function $\Theta(t)$. Those graphs that contain an internal curly line are 
exactly the one-particle reducible (1PR) diagrams.

\subsubsection{\label{sec:mf:pert:consistency}%
Regularity of the perturbative series%
}
The perturbation expansion leaves some questions, that have to be addressed.
It is not immediately obvious, that taking the disorder average of (\ref{eq:zeta2})
gives the result in (\ref{eq:diszeta2}), i.e. $\diszeta_2(t)=\disav{\zeta_2(t)}$.
However, a short calculation, using integration by parts reveals this relation to hold.

Another, much deeper problem is related to the diagrams involving a curly line in 
their interior.
Due to the curly line, they grow linearly in time. Already in section
\ref{sec:walls:highord} and appendix \ref{app:dg4proof}, we 
have mentioned that there are
trees that contain lines the assigned momentum of which equals 0.
Here, for the mean-field problem these lines are found as the troublesome 
curly lines: they connect a subtree with internal Gau\ss ian pairing.
Koplik and
Levine \cite{KL:PRB85} explicitly checked for a time independent driving $h(t)=h$
up to sixth
order, that the problematic terms of the 1PR diagrams mutually cancel. 
We give a very general version of this proof, 
that holds for any time-dependence of the
driving field $h(t)$ and covers all perturbative
orders. To illustrate, how this works, we present the calculation for the
fourth order here. The somewhat technical induction step, which extends our argument
to all orders is given in appendix \ref{app:reg}.
For simplicity, we work with the velocity diagrams, that are obtained by just
removing the curly line from the root.
\begin{widetext}
\begin{align}
2\cdot\>\parbox{16mm}{\includegraphics{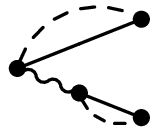}}=&\int\limits_0^t\cd t_1\>\e^{-c(t-t_1)}
\Delta''[Z(t)-Z(t_1)]\int\limits_0^t\cd t_2\int\limits_0^{t_2}\cd t_3\>
\e^{-c(t_2-t_3)}\Delta'[Z(t_2)-Z(t_3)]\\
\parbox{24mm}{\includegraphics{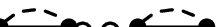}}=&\int\limits_0^t\cd t_1\>\e^{-c(t-t_1)}
(-\Delta''[Z(t)-Z(t_1)])\int\limits_0^{t_1}\cd t_2\int\limits_0^{t_2}\cd t_3\>
\e^{-c(t_2-t_3)}\Delta'[Z(t_2)-Z(t_3)]\\
=&-2\cdot\>\parbox{16mm}{\includegraphics{vdiag1.eps}}+S\\
S=&\int\limits_0^t\cd t_1\int\limits_0^{t_1}\cd t_2\>\e^{-c(t-t_2)}
\Delta''[Z(t)-Z(t_2)]\int\limits_0^{t_1}\cd t_3\>
\e^{-c(t_1-t_3)}\Delta'[Z(t_1)-Z(t_3)]
\end{align}
\end{widetext}
The modification of the second diagram to express it as the sum of the first and $S$
is merely integration by parts for the integral over $t_1$.
The term $S$ now corresponds to the sum of the two diagrams. It is easy to see, that $S$
remains bounded for large times. Every time integral carries an exponential
damping term.
Basically, we have thereby established, that the
perturbation series exists and is well-behaved in the sense, that there are
no terms that lead to an overall unbounded growth in time.

\subsubsection{\label{sec:mf:pert:valpert}%
Validity of perturbation theory%
}
Still, the question is open, whether one may assume
$\zeta$ to be small compared to $\ell$. This was a requirement for
the Taylor expansion (\ref{eq:mfdisexpansion}) to be valid.
If $c$ is large, any particle moving in a particular realisation of a
disorder potential is strongly bound to the disorder averaged position.
This prevents it from exploring the own disorder environment and thus
large $c$ effectively scale down $\eta$. All realisations stay close to the
disorder averaged position, the mean deviation being approximated by 
$\eta/c$. 
A problem now occurs, if the disorder averaged
position deviates strongly from the $\eta=0$ solution. For $h\gg\eta$ this can
only happen during those periods, where $h(t)$ takes on small values.
The time that has to elapse until every system has adopted to its
own disorder realisation, and hence the time until the system can be pinned,
is $c^{-1}$ (see below). For perturbation theory to work, 
this time must be large compared to the length of the period
during which $h\le\eta$, which we roughly estimate as $\eta/(\omega h)$. This
gives us a second condition for the applicability of perturbation theory:
$h/\eta\gg c/\omega$.

In summary, the conditions for perturbation theory to hold are the
following. The driving force amplitude $h$ has to be large compared to 
$\eta$, $h/\eta\gg\max\{c/\omega,1\}$ to make the series 
expansion work and
to guarantee that the disorder averaged solution stays close to the
$\eta=0$ trajectory (around which we expand). 
Moreover, $c$ must be large ($c\gg\eta/\ell$)
to ensure proximity of each realisation to the disorder average.

\begin{figure}[t]
\includegraphics[width=\columnwidth]{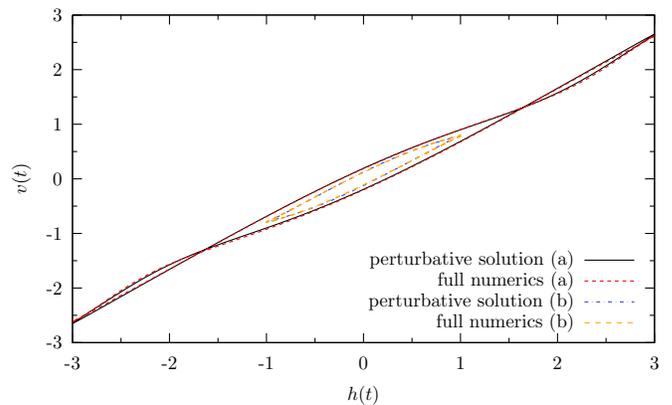}
\caption{\label{fig:stvollvgl} Comparison of the full numerical
solution of equation (\ref{eq:mfgeom}) with the result obtained
from the first non-vanishing perturbative order eq. (\ref{eq:v1stop})
for
(a) $h=3.0$, $c=3.0$, $\eta=1.5$ and 
(b) $h=1.0$, $c=1.0$, $\eta=0.6$.
The units of $t$ and $z$ are chosen such that
$\omega=\ell=1$.}
\end{figure}%
The diagrammatic prescription
yields up to terms of order ${\cal O}(\eta^2)$
\begin{equation}
\label{eq:v1stop}
v(t)=h\cos\omega t+\eta^2\int\limits_0^t\cd t'\>
\e^{-c(t-t')}\Delta'[Z(t)-Z(t')].
\end{equation}
A direct
comparison of the numerical solution of (\ref{eq:v1stop}) and
the full equation of motion (\ref{eq:mfgeom}), 
shown in fig. \ref{fig:stvollvgl}, confirms an excellent 
agreement.

\subsection{\label{sec:ac:harmon}%
Perturbative harmonic expansion%
}

For an ac driving force, even the lowest perturbative order for the
velocity, equation (\ref{eq:v1stop}) is a very complicated expression.
We know from the numerics that, in the stationary state, the velocity is
given by a periodic function with periodicity $\omega^{-1}$. This
recommends to aim a harmonic expansion of the mean velocity $v$, i.e.
to ask for the Fourier coefficients $a_N$ and $b_N$ in the ansatz
\begin{align}
v(t)=\sum\limits_{N=1}^\infty\big[a_N\cos N\omega t+b_N\sin N\omega t\big],
\end{align}
It is now an important question, 
whether the first-order perturbative correction can further be simplified
to get an analytic description of interesting features of the trajectory.
One idea could be, to take only the lowest Fourier modes. In this section,
we want to investigate, under which circumstances this could be possible.

Starting from the first order result for $v$, given by eq. (\ref{eq:v1stop}),
we express the disorder correlator by its Fourier transform
\begin{align}
\Delta'[Z(t)-Z(t')]=\int\limits{\cd q\over 2\pi}(\ii q)\Delta(q)
\e^{\ii q{h\over\omega}[\sin\omega t-\sin\omega t']}
\end{align}
and expand the exponential term in a double Fourier series
in $t$ and $t'$, respectively:
\begin{eqnarray*}
\e^{\ii a\sin\omega t}&=&\sum\limits_{n=-\infty}^\infty
J_n(a)\e^{\ii n\omega t}\\
\int\limits_0^t\cd t'\>\e^{-c(t-t')-\ii a\sin\omega t'}&=&
\sum\limits_{n=-\infty}^\infty
J_n(-a){\e^{\ii n\omega t}-\e^{-ct}\over c+\ii n\omega}.
\end{eqnarray*}
Here, $J_n(a)$ are the Bessel functions of the first kind.
As we are interested only in the behaviour for large enough times (the steady state
solution),
we remove all terms that are damped out exponentially for $t\gg c^{-1}$ 
from the very beginning. Note,
that $c^{-1}$ is indeed the time scale for the transience, as has been claimed before.

For the mean velocity, we obtain
\begin{widetext}
\begin{align}
\label{eq:mffourexp}
v(t)=h\cos\omega t+\eta^2\sum\limits_{m,n=-\infty}^\infty
\int{\cd q\over2\pi}(\ii q)\Delta(q)J_m\left(q{h\over\omega}\right)
J_n\left(-q{h\over\omega}\right)
{\e^{\ii(m+n)\omega t}(c-\ii n\omega)\over c^2+n^2\omega^2}.
\end{align}
\end{widetext}
In principle, this is already a Fourier series representation, not very
elegant, though. The argument $(m+n)\omega t$ of the expansion basis 
exponentials promises a rather complicated structure for the coefficients.
A first observation, however, can already be made: Under the $q$ integral we
find an odd function $(\ii q)\Delta(q)$ and a product of two Bessel functions
of order $m$ and $n$, respectively. For the $q$-integral to result in a finite value, 
a function is required that is not odd in $q$. This necessitates the product
of the two Bessel functions to be odd, or, equivalently, $m+n$ to be an odd number.
Whence, we conclude, that to first perturbative order, our symmetry argument
(Fourier coefficients for even $N$ must vanish) is fulfilled exactly.

It requires some
tedious algebra to collect all contributions belonging to a
certain harmonic order from the double series. Eventually, we obtain
a series expansion
\begin{align}
\label{eq:mf1sto}
{{v}(t)\over h}=\cos\omega t+
\sum\limits_{N=1}^\infty\bigg[&
A_N\left({\eta\over h},{\omega\over c},{h\over\omega\ell}\right)\cos N\omega t+\nonumber\\
&B_N\left({\eta\over h},{\omega\over c},{h\over\omega\ell}\right)\sin N\omega t
\bigg].
\end{align}

Note, that taking $\omega\to 0$ is forbidden here, as we used $\omega\ne 0$ while
deriving the coefficients and moreover perturbation theory breaks down (recall that
$h/\eta\gg c/\omega$). The same holds for $\ell\to 0$.
The remaining extreme limits $\omega\to\infty$ and 
$\ell\to \infty$ are not interesting, since in these limits the disorder is rendered 
unimportant. Therefore, in the following, we assume finite (positive) values for
$\ell$ and $\omega$ and moreover set them equal to one $\omega=\ell=1$,
by appropriately choosing the units for $z$ and $t$.

Now, we are left with three dimensionless parameters: 
$h$, $c$ and $\eta$.
The dependence of the first order perturbative Fourier coefficients on $\eta$ is trivial. 
The dependence on $c$
is also evident, as can be read off from (\ref{eq:mffourexp}). For larger $c$,
the system is more tightly bound to the non-disordered solution, 
supressing perturbative corrections.

\begin{figure}
\includegraphics[width=\columnwidth]{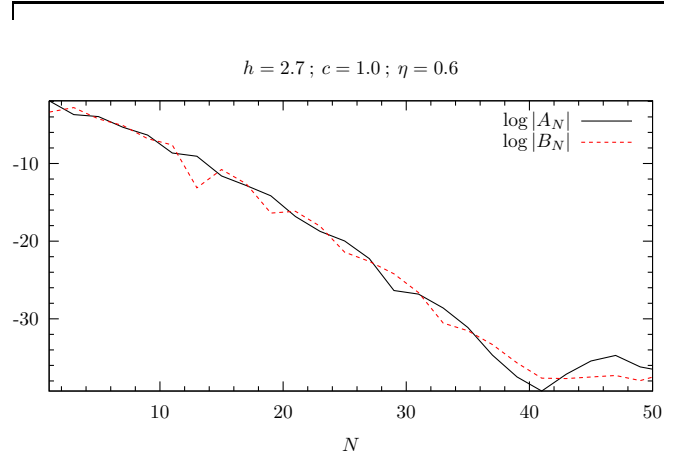}
\caption{\label{fig:habh1}%
Plotting the logarithms of $|A_N|$ and $|B_N|$ reveals
the exponential decay with $N$. In the regime where numerical errors do not
dominate the result, a linear regression seems appropriate.
}%
\end{figure}%
\begin{figure}
\includegraphics[width=\columnwidth]{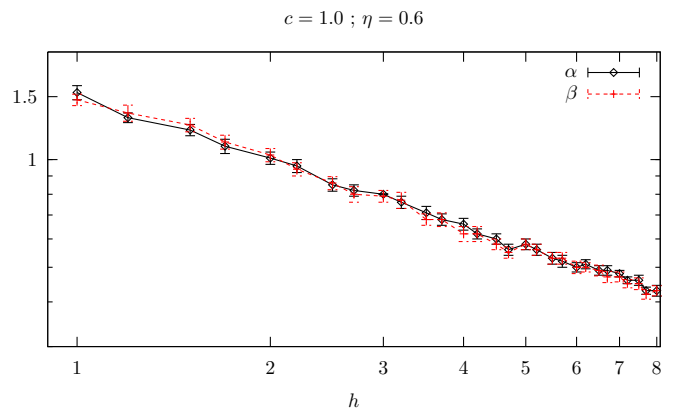}
\caption{\label{fig:habh2}%
Performing the
linear regression for many $h$ yields slopes $\alpha$ and $\beta$ appearing to depend 
on $h$ in a power-law fashion.}
\end{figure}%
The most interesting but also the most difficult is the dependence of the Fourier coefficients
on $h$. Actually, there are two competing effects. On the one hand, large driving strengths
render the disorder unimportant in all cases accessible through perturbative methods.
On the other hand, 
if one thinks of $g(Z(t))$ as a function of time, the more rapid $Z(t)$ 
changes the more $g$ fluctuates on short time scales and thus
brings higher frequency contributions to $v(t)$. The first remark is reflected in
the overall weight of the Fourier coefficients as corrections to the
non-disordered case, decreasing with $h$. The second idea is expected to express itself 
in the decay
of the Fourier coefficients with $N$. The larger $h$, the weaker we expect this 
decay to be.

The dependence of the higher harmonics on $h$ is hidden in the Bessel functions
in (\ref{eq:mffourexp}).
The first extremum of the Bessel functions
shifts to larger values as $m$ and $n$ increase, respectively. 
However, the complicated way in which these Bessel functions enter
$A_N$ and $B_N$ hinders an analytic access to the decay law.
A numerical determination of the Fourier coefficients for the perturbative result
reveals an exponential decay, as shown in fig. \ref{fig:habh1}. The noisy behaviour for 
$N\ge 40$ is due to numerical fluctuations. Note, that these fluctuations are of
the order $10^{-14}$, which is quite reasonable. The plot in fig. \ref{fig:habh1}
is mere illustration of a more general phenomenon. This exponential decay
has been found for many sets of parameters, thus one is led to the ansatz
\begin{equation}
|A_N|\sim{\eta^2\over h^2}\e^{-\alpha N}\quad;\quad 
|B_N|\sim{\eta^2\over h^2}\e^{-\beta N},
\end{equation}
where $\alpha$ and $\beta$ can be estimated through a linear regression up to a suitable 
$N_{\text{max}}$. Of course, it is not expected, that $\alpha$ and $\beta$ are distinct,
nor that they depend on the parameters in different ways. Determining both just doubles
the amount of available data.

As our results are first-order perturbative, $\alpha$ and $\beta$ must not 
depend on $\eta$. The main interest now focusses
on the dependence of the decay constants on $h$. The results from a linear regression for
a series of $h$-values, $c$ and $\eta$ kept fixed, suggest a power-law dependence
\begin{equation}
\label{eq:regression}
\alpha(h,c)=C_\alpha(c)\cdot h^{-\xi_\alpha},\quad\beta(h,c)=C_\beta(c)\cdot h^{-\xi_\beta}.
\end{equation}
Fig. \ref{fig:habh2} displays this relation for a particular example. Repeating
this data collection and subsequent regression for different values for $c$ and $\eta$
yields the results summarised in table \ref{tab:habh}. 
\begin{table}
\caption{\label{tab:habh}Results for the regression (\ref{eq:regression}).}
\begin{ruledtabular}
\begin{tabular}{cccccc}
$c$&$\eta$&$C_\alpha$&$C_\beta$&$\xi_\alpha$&$\xi_\beta$\\
\hline
1.0&0.6&1.52&1.52&0.61&0.61\\
1.5&0.6&1.56&1.56&0.58&0.59\\
2.0&0.6&1.56&1.59&0.58&0.60\\
2.5&0.6&1.63&1.62&0.61&0.61\\
3.0&1.0&1.66&1.66&0.63&0.63\\
3.5&1.0&1.71&1.68&0.62&0.62\\
4.0&1.0&1.69&1.65&0.61&0.60\\
4.5&1.0&1.72&1.68&0.62&0.62\\
5.0&2.0&1.71&1.67&0.61&0.60\\
5.5&2.0&1.73&1.73&0.61&0.62\\
6.0&2.0&1.77&1.72&0.62&0.62\\
6.5&3.0&1.76&1.77&0.62&0.62
\end{tabular}
\end{ruledtabular}
\end{table}
While the exponent $\xi$ appears constant $\xi\approx 0.6$, the prefactor
seems to depend on $c$. 
\begin{figure}
\includegraphics[width=\columnwidth]{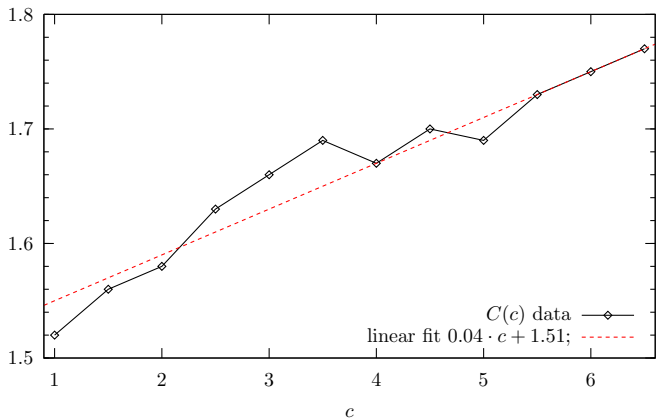}
\caption{\label{fig:vorfc}Plot of the change of the prefactor $C(c)=(C_\alpha+C_\beta)/2$ 
in (\ref{eq:regression})
on $c$. The linear fit yields a fairly tiny slope.}
\end{figure}%
An attempt to redo the same procedure, done for $h$,
with the parameter $c$ to gain information
about the functional dependence of $\alpha$ and $\beta$ on $c$ yields a complicated
but rather weak dependence, which gives no further insight. The linear fit
in fig. \ref{fig:vorfc} gives a fairly tiny slope, so the dependence of the decay
constants on $c$ may be assumed to be weak.

Certainly, it is desirable to ascertain
the validity of this decay law beyond perturbation theory. In a few words, it ought to
be explained, why we have not been able to do it. First of all, the logarithmic plots
of the Fourier coefficients in fig. \ref{fig:habh1} exhibit fluctuations around the
linear decrease. This ``noise'' is authentic and not attributed to numerical
inaccuracies. The exponential decay of the Fourier coefficients is superimposed on
a true, complicated dependence. Hence, it requires a lot of data points to obtain
reasonable data. Since the Fourier coefficients for even $N$ vanish, 
in the example of fig. \ref{fig:habh1} the
regression can be carried out over around 15-20 data points. This is a fair number. The
quality relies heavily on the accuracy of the numerical determination of the
Fourier coefficients. In Fourier analyses of the numerics for the full equation
of motion (\ref{eq:mfgeom}), we did not manage to get a precision better than
of the order of $10^{-3}$. This means, the regression has to be stopped at 
$N_{\text{max}}$, where $\log A_{N_{\text{max}}}\approx -7$. In the example
of fig. \ref{fig:habh1}, this leaves us with less than 5 data points. In view of
the natural fluctuations, a linear regression is not sensible any more.

In summary, we have numerically established the dependence of the decay constans for 
the Fourier coefficients on $h$: $\alpha,\beta(h)=C(c)\cdot h^{-\xi}$
with $\xi\simeq 0.6$.

\section{\label{sec:conclusions}%
Conclusions%
}

We have seen that the perturbation expansion
for ac-driven elastic interfaces
in random media fails for interface dimensions $D\le 4$.
We have resolved this puzzle,
and the reason for the strange behaviour of perturbation theory
has been found to be connected with
the disorder-dependent time scale $\tau$ 
(cf. (\ref{eq:tautscale})) which measures the time for the initial
roughening of interfaces in random environments. 
Due to its appearence within the disorder
averaged velocity $v$ as a transience relaxation time, 
attemps to determine $v$ via a perturbation expansion in the strength
of the disorder entail terms of unbounded growth in time. 
Although, most probably, the expansion
yields the true solution if it could be summed up, it is useless, because
the finite perturbative orders grow in powers of time. They offer a good
approximation to the full solution only on time scales small
compared to the transience relaxation time and fail to
describe correctly the behaviour on large time scales.
On the other hand, as we have signified, the perturbation expansion
for $D>4$ is regular. 

Therefore, theoretical work towards an understanding of the
important problem of ac-driven interfaces exposed to disorder
has to take the route via the related mean-field problem. 
The mean-field equation of motion formally corresponds to
$D=\infty$ and admits a regular
perturbative treatment that, where applicable, agrees very well
with the numerics for the full equation of motion. It has been shown,
that non-regular diagrammatic contributions 
cancel among each other, leaving a well-behaved perturbative
expansion.

Further, the solutions to the mean-field equation of motion
have many features in common with the problem in finite
dimension. Therefore, they
can be useful to study some properties of the original problem,
like the velocity-hysteretis.

The mean-field perturbation expansion helped to
improve numerical results, which allowed us to
establish the dependence of the decay constants of the Fourier
modes on $h$ as a power law.

\begin{acknowledgments}
For fruitful discussions and the introduction into the problem 
I am grateful
to T. Nattermann. Further, I want to thank G. M. Falco, 
A. Fedorenko, A. Glatz, A. Petkovi\'c and
Z. Ristivojevic for discussions.
Finally, I would like to acknowledge financial support by 
Sonderforschungsbereich 608.
\end{acknowledgments}

\appendix

\section{\label{app:fiappr}%
The functional integral approach%
}
Apart from a perturbative expansion of the equation of motion (\ref{eq:weom}), 
it is also possible to compute the disorder averaged correlation functions
via a functional integral approach \cite{MartinSiggiaRose:PRA73,DeDominics:PRB78}.
Starting from (\ref{eq:weom})
\begin{align}
\partial_tz_{x,t}=\Gamma\nabla_x^2z_{x,t}+h(t)+
u\cdot g(x,z_{x,t})\equiv F[z_{x,t}]
\end{align}
(with $z_{x,t}$ being a short-hand notation for $z(x,t)$),
we note that
\begin{align}
\disav{A[z_{x,t}]}&=\disav{\int\cD z_{x,t}\>A[z_{x,t}]\,\delta\Big(
\partial_tz_{x,t}-F[z_{x,t}]\Big)}
\end{align}
for any functional $A[z_{x,t}]$. Writing
\begin{align}
\delta\Big(
\partial_t&z_{x,t}-F[z_{x,t}]\Big)
\nonumber\\
&=\int\cD\hat z_{x,t}\>
\exp\left[\ii\int_{x,t}\hat z_{x,t}(\partial_tz_{x,t}-F[z_{x,t}])\right]
\end{align}
and using the usual cumulant expansion for Gau\ss ian disorder,
we find
\begin{align}
\disav{A[z_{x,t}]}&={1\over\cZ}\int\cD\hat z_{x,t}\cD z_{x,t}\>
A[z_{x,t}]\,\e^{\cS[\hat z,z]}
\\
\cZ&=\int\cD\hat z_{x,t}\cD z_{x,t}\>\e^{\cS[\hat z,z]}
\\
\cS[\hat z,z]&=\cS_0[\hat z,z]+\cS_h[\hat z,z]+\cS_\dis[\hat z,z].
\end{align}
The three contributions to the action for this functional integral formula
read
\begin{align}
\cS_0[\hat z,z]&=\ii\int_{x,t}\hat z_{x,t}\big[\partial_t-\Gamma\nabla_x^2\big]
z_{x,t}
\\
\cS_h[\hat z,z]&=\ii\int_{x,t}\hat z_{x,t}h(t)
\\
\cS_\dis[\hat z,z]&=-{u^2\over 2}\int_{x,t,t'}\hat z_{x,t}\hat z_{x,t'}
\Delta[z_{x,t}-z_{x,t'}].
\end{align}
Certainly, we can also compute correlation functions $A[z_{x,t},\hat z_{x',t'}]$.
For example, we re-obtain the propagator (\ref{eq:pertprop}) by calculating the
response function
\begin{align}
\label{eq:fia:prop}
\ii\disav{z_{x,t}\hat z_{x',t'}}_{\cS_0}=G(x-x',t-t').
\end{align}
To set up a perturbative expansion in $u$, we decompose in the same way as in
section \ref{sec:walls:perturbation} $z_{x,t}=Z_t+\zeta_{x,t}$ with $Z_t=(h/\omega)
\sin\omega t$.
Instead of $z_{x,t}$ we now have to consider the functional integral over $\zeta_{x,t}$.
Now
\begin{align}
\cS[\hat z,\zeta]&=\cS_0[\hat z,\zeta]+\cS_\dis[\hat z,\zeta]
\\
\cS_\dis[\hat z,\zeta]
&=-{u^2\over 2}\int_{x,t,t'}\hat z_{x,t}\hat z_{x,t'}
\Delta[Z_t+\zeta_{x,t}-Z_{t'}-\zeta_{x,t'}].
\end{align}
To compute correlation functions perturbatively in powers of $u$ requires
to expand the normalisation $\cZ$ as well. A small reflection shows that
the lowest order contribution coming from $\cZ$ is of order $\order{u^4}$.
Thus, if we want to calculate the velocity to order $\order{u^2}$ we can ignore
the $u$-dependence of $\cZ$ and write
\begin{align}
\label{eq:fia:v}
v(t)&=h(t)+\partial_t\disav{\zeta_{x,t}}
\nonumber\\
&=h(t)+\partial_t\disav{\zeta_{x,t}\cS_\dis[\hat z,\zeta]}_{\cS_0}+
\order{u^4}.
\end{align}
Of course, we may only retain $\cS_\dis[\hat z,\zeta]$ up to terms of order
$\order{u^2}$, i.e.
\begin{widetext}
\begin{align}
\cS_\dis[\hat z,\zeta]
&=-{u^2\over 2}\int_{x,t,t'}\hat z_{x,t}\hat z_{x,t'}\big[\Delta[Z_t-Z_{t'}]
+\Delta'[Z_t-Z_{t'}](\zeta_{x,t}-\zeta_{x,t'})\big]
+\order{u^4}.
\end{align}
Averages with respect to $\cS_0$ are Gau\ss ian, thus Wick's theorem applies and 
with (\ref{eq:fia:prop}) we recover the result from equation (\ref{eq:v2ndo})
\begin{align}
v(t)&=h(t)-{u^2\over 2}\partial_t\int_{x',t_1,t_2}\Delta'[Z_{t_1}-Z_{t_2}]\left\{
\disav{\zeta_{x,t}\hat z_{x',t_1}}_{\cS_0}\disav{\zeta_{x,t_1}\hat z_{x',t_2}}_{\cS_0}-
\disav{\zeta_{x,t}\hat z_{x',t_2}}_{\cS_0}\disav{\zeta_{x,t_2}\hat z_{x',t_1}}_{\cS_0}
\right\}+\order{u^4}
\nonumber\\
&=h(t)-u^2\partial_t\int_{x',t_1,t_2}\Delta'[Z_{t_1}-Z_{t_2}]\{
-\ii G(x-x',t-t_1)\}\{-\ii G(x-x',t_1-t_2)]\}+\order{u^4}
\nonumber\\
&=h(t)+u^2\int_{t'}\Delta'[Z_{t}-Z_{t'}]G(0,t-t')+\order{u^4}.
\end{align}
\end{widetext}
It is clear, that we can continue equation (\ref{eq:fia:v}) to higher orders in
$u$ by taking into account higher orders in $u$ of $\exp(\cS_\dis)$ 
as well as higher order corrections from $\cZ$.
This way, it is possible to rearrive at the graphical expansion presented
in section \ref{sec:walls:highord}, albeit along a little more complicated route.

\section{\label{app:dg4proof}%
The regularity of all perturbative orders
in case $D>4$%
}
Let us start with an example graph at which we demonstrate the steps, that
are then generalised further down. We
consider the following fourth order contribution to the correction of
the disorder averaged velocity
\begin{widetext}
\begin{align}
\parbox{25mm}{\includegraphics{exdiag1.eps}}=&
\int\limits_0^t\cd t_1\int\limits_0^{t_1}\cd t_2\int\limits_0^{t_2}\cd t_3
\prod\limits_{i=1}^3\int\cd^Dx_i\nonumber\\
&G(x-x_1,t-t_1)G(x_1-x_2,t_1-t_2)G(x_2-x_3,t_2-t_3)\times\nonumber\\
&\delta(x-x_2)\delta(x_1-x_3)(-\Delta''[Z-Z_2])\Delta'[Z_1-Z_3]\\
\label{eq:bspdiagvereinf}
=&\int\limits_0^t\cd t_1\int\limits_0^{t_1}\cd t_2\int\limits_0^{t_2}\cd t_3\>
(-\Delta''[Z-Z_2])\Delta'[Z_1-Z_3]\times\nonumber\\
&\int\cd^Dx'\>G(x-x',t-t_1)G(x'-x,t_1-t_2)G(x-x',t_2-t_3)\\
\label{eq:bspintox}
\int\cd^Dx'G(x-x',t-t_1)&G(x'-x,t_1-t_2)G(x-x',t_2-t_3)=\nonumber\\
&\int{\cd^Dk_1\cd^Dk_2\over (2\pi)^{2D}}
\e^{-\Gamma\big[k_1^2(t-t_1)+(k_1+k_2)^2(t_1-t_2)+k_2^2(t_2-t_3)\big]}
\end{align}
\end{widetext}
We can now assign momenta to the branches of the tree
\begin{align}
\parbox{25mm}{\includegraphics{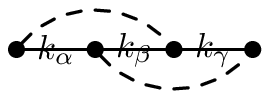}},
\end{align}
and we can read off from (\ref{eq:bspintox}), that
\begin{align}
k_\alpha=k_1\,,\quad
k_\beta=k_1+k_2\,,\quad
k_\gamma=k_2.
\end{align}
Obviously, $k_1$ describes the momentum ``flowing'' from the root to vertex 2
and $k_2$ is the momentum ``flow'' from vertex 1 to vertex 3.
Estimating the disorder correlator derivatives in equation
(\ref{eq:bspdiagvereinf}) by constants (that we can
drop since they do not influence the asymptotics $t\to\infty$), all that
has to be done is the integration of the right-hand side of
(\ref{eq:bspintox}) over the 3 time variables $t_1$, $t_2$ and $t_3$.
We start with the outermost leaf, $t_3$, and have
\begin{align}
\label{eq:tintest}
\int\limits_0^{t_2}\cd t_3\>\e^{-\Gamma k_2^2(t_2-t_3)}=
{1\over \Gamma k_2^2}\left[1-\e^{-\Gamma k_2^2t_2}\right]
\le{1\over\Gamma k_2^2}.
\end{align}
Proceeding in the same manner for the remaining time variables, we finally find
\begin{align}
\bigg|\>\parbox{25mm}{\includegraphics{exdiag1.eps}}\bigg|\le&
\int{\cd^Dk_1\cd^Dk_2\over (2\pi)^{2D}}{1\over\Gamma^3 k_1^2(k_1+k_2)^2k_2^2}.
\end{align}
This is certainly finite for $D>4$.

Now, we turn to the general argument for all trees of general (even) order $2q$.
Let ${\cal T}_q$
denote the set of all rooted trees with $q$ vertices, and ${\cal P}(T)$ all possible
unordered pairings of vertices of $T\in{\cal T}_q$.
Let moreover $B_T$ be the set of all branches
of the tree $T$. We want to agree that a branch $b=(b_1,b_2)$ has $b_1$ always closer
to the root. Then, we have in general for every order
\begin{align}
\disav{v_{2q}}=\sum\limits_{T\in{\cal T}_{2q}}\sum\limits_{P\in{\cal P}(T)} A_{T,P}.
\end{align}
Here, $A_{T,P}$ is a single diagram, namely a tree $T$ where all pairs $(p_1,p_2)\in P$
are connected by dashed lines for the Gau\ss ian disorder average. The disorder
correlators that enter $A_{T,P}$ can be estimated by constants
$|\Delta^{(m)}[Z'-Z'']|\le c_m\ell^{-m}$, that we drop in the following. They are finite
and do not influence the behaviour of $A_{T,P}$ as $t\to\infty$.
Denoting the root vertex by $r$, we estimate
\begin{widetext}
\begin{align}
|A_{T,P}|\le&\bigg(\prod\limits_{b\in B_T}\int{\cd^Dk_b\over(2\pi)^D}\int\limits_0^{t_{b_1}}
\cd t_{b_2}\int\cd^Dx_{b_2}\bigg)\bigg(\prod\limits_{b\in B_T}
\e^{-\Gamma k_b^2(t_{b_1}-t_{b_2})+\ii k_b(x_{b_1}-x_{b_2})}\bigg)
\prod\limits_{p\in P}\delta(x_{p_1}-x_{p_2})\\
=&\bigg(\prod\limits_{b\in B_T}\int{\cd^Dk_b\over(2\pi)^D}\int\limits_0^{t_{b_1}}
\cd t_{b_2}\bigg)\bigg(\prod\limits_{p\in P\atop r\notin p}\int\cd^Dx_{p_1}\bigg)
\exp\bigg[-\Gamma\sum\limits_{b\in B_T}k_b^2(t_{b_1}-t_{b_2})\nonumber\\
&+\ii\sum\limits_{p\in P}x_{p_1}\sum\limits_{b\in B_T}k_b
\left(\delta_{b_1,p_1}+\delta_{b_1,p_2}-\delta_{b_2,p_1}-\delta_{b_2,p_2}\right)
\bigg]
\end{align}
The integration over the remaining $x$-coordinates brings delta-functions
for the momenta:
\begin{align}
\label{eq:atpintres}
|A_{T,P}|\le&\bigg(\prod\limits_{b\in B_T}\int{\cd^Dk_b\over(2\pi)^D}\int\limits_0^{t_{b_1}}
\cd t_{b_2}\bigg)\exp\bigg[-\Gamma\sum\limits_{b\in B_T}k_b^2(t_{b_1}-t_{b_2})\bigg]
\nonumber\\
&\prod\limits_{p\in P\atop r\notin p}(2\pi)^D\delta\bigg(\sum\limits_{b\in B_T}
k_b\left(\delta_{b_1,p_1}+\delta_{b_1,p_2}-\delta_{b_2,p_1}-\delta_{b_2,p_2}\right)
\bigg)
\end{align}
\end{widetext}
The $q-1$ delta-functions for the momenta mean, that the net out-flow (away from the root)
of momentum from a vertex $p_1$ equals the net in-flow of momentum for the 
Gau\ss ian partner
vertex $p_2$ (i.e. $(p_1,p_2)\in P$). There is no delta-function ensuring this
for the root and its partner vertex. However, this is not needed. The root itself has
only out-flow of momentum and that has to be absorbed by its partner vertex
to ensure the balance for all other pairs. This explains, why no
exponential function involving $x$ appears any more in (\ref{eq:atpintres}). 
As a result, we can assign
a momentum $k_p$ to each pair $P\ni p=(p_1,p_2)$, describing the net momentum 
transfer from the vertex $p_1$ to $p_2$.
This insight advises to choose the momentum associated to the $q$ Gau\ss ian pairs
as the $q$ integration variables $k_p$.
The momentum assigned to a bond
$k_b$ denotes the total flow of momentum through this bond, determined by the
source or drain properties of the bond boundary vertices $b_1$ and $b_2$.
The rest is now easy, if we follow
the estimate for the time integrals from the equation (\ref{eq:tintest}) in the
example before. We end up with
\begin{align}
\label{eq:atpest}
|A_{T,P}|\le&\bigg(\prod\limits_{p\in P}\int{\cd^Dk_p\over(2\pi)^D}\bigg)
\prod\limits_{b\in B_T}{1\over\Gamma k_b^2}.
\end{align}
One first puzzle related to (\ref{eq:atpest}) is obvious: 
there may be bonds, that carry momentum $k_b=0$.
This exactly happens for lines that connect to a subtree which has only internal
Gau\ss ian pairings, i.e. the whole momentum flow remains inside the subtree.
Such lines are problematic in all dimensions, even in the mean-field theory,
where they correspond to the curly lines in the graphical expansion for the
mean-field equation of motion (cf. section \ref{sec:mf:pert}). 
The resolution of this problem by cancellations among several such diagrams 
is technically a little easier for the mean-field case,
where we performed it (cf. section \ref{sec:mf:pert:consistency}
and appendix \ref{app:reg}). The idea
of the mean-field proof exactly applies here as well, just
in mean-field 
we do not carry the load of momentum integrals. The adaption of the proof itself is
straightforward.

Another problem in connection with (\ref{eq:atpest}) is the fact, that we have
only $q$ momenta to integrate over but $2q-1$ bonds. So for every given order it
happens for a couple of diagrams, that some momentum, $k_p$ say, 
appears to the power $k_p^{-2q}$ in 
(\ref{eq:atpest}). Thus, the integration over $k_p$ has infrared problems in
$D\le 2q$, but integration over the other momenta works in $D>2$. Actually, this problem 
can again be resolved by summing several such diagrams in multiple steps.
After the first step, one obtains a result, that has $k_p$ divergent in $D\le 2q-2$
but another momentum, $k_{p'}$ say, divergent in $D\le 4$ instead of $D\le 2$.
So, viewed as a function of $k_p$ the intermediate result is better behaved, but worse
for $k_{p'}$. Let us take a closer look at this.
We start with a glimpse on this issue for the expansion of $\disav{v_6}$, 
where we have checked that everything remains bounded in $D>4$.
The expansion of $\disav{v_6}$ contains, among others, the following diagrams
\begin{align}
\label{eq:6od1}
D_1^6&=\parbox{25mm}{\includegraphics{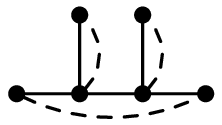}}\\ \nonumber\\
\label{eq:6od2}
D_2^6&=\parbox{33mm}{\includegraphics{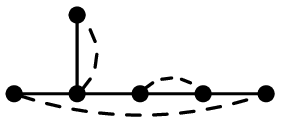}}\\ \nonumber\\ 
\label{eq:6od3}
D_3^6&=\parbox{33mm}{\includegraphics{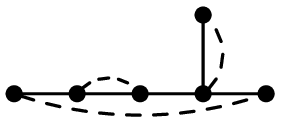}}\\ \nonumber\\
\label{eq:6od4}
D_4^6&=\parbox{41mm}{\includegraphics{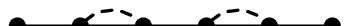}}.
\end{align}
With regard to their topological structure, 
these are the only diagrams that have problems in $D>4$, other
such diagrams are constructed from them by a different choice of the root. 
Incidentally, $D_2^6$ and $D_3^6$ are topologically equivalent as well.
We can easily see that in each diagram $D_1^6,\ldots,D_4^6$, 
there are 3 bonds that
carry the momentum $k_1$ belonging to the pair of the root and the outermost
vertex. The momenta associated to the other two Gau\ss ian pairs, $k_2$
and $k_3$, occur only within exactly one bond. Thus, according to
equation (\ref{eq:atpest}) the integration over $k_1$
produces problems in $D\le 6$, whereas integration over $k_2$ and $k_3$ 
is harmless in $D>2$.
One can now check, 
that the sum $4D^6_1+2D^6_2$ as well as the sum of $2D^6_3+D^6_4$ is regular in $D>4$.
The factors that I have put count the incidence of each diagram
in the expansion.

The mechanism of combining diagrams to improve the properties with respect
to integration over some $k_i$ and make it worse for some other $k_j$ can already 
be demonstrated at graphs of order $4$. 
If, from $D^6_1$ and $D^6_2$ the root and the upper vertex closest 
to the root are removed and the outermost leaf connected to the new root, we have
the two diagrams of fourth order (cf. (\ref{eq:bentw4}))
\begin{widetext}
\begin{align}
\label{eq:4od1}
2D_1^4&=2\cdot\>\parbox{17mm}{\includegraphics{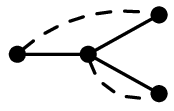}}\nonumber\\
&=
\int_{k_1,k_2}\int\limits_0^t\cd t_1\>\e^{-\Gamma k_1^2(t-t_1)}
\int\limits_0^{t_1}\cd t_2\>\e^{-\Gamma k_2^2(t_1-t_2)}\Delta''[Z_1-Z_2]
\int\limits_0^{t_1}\cd t_3\>\e^{-\Gamma k_1^2(t_1-t_3)}\Delta'[Z-Z_3]\\
\label{eq:4od2}
D_2^4&=\>\parbox{25mm}{\includegraphics{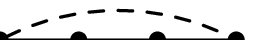}}\nonumber\\
&=
-\int_{k_1,k_2}\int\limits_0^t\cd t_1\>\e^{-\Gamma k_1^2(t-t_1)}
\int\limits_0^{t_1}\cd t_2\>\e^{-\Gamma k_2^2(t_1-t_2)}\Delta''[Z_1-Z_2]
\int\limits_0^{t_2}\cd t_3\>\e^{-\Gamma k_1^2(t_2-t_3)}\Delta'[Z-Z_3]\\
&=-2\cdot\>\parbox{17mm}{\includegraphics{4odiag1.eps}}+S
\end{align}
\begin{align}
\label{eq:4ospi}
S&=\int_{k_1,k_2}\int\limits_0^t\cd t_1\>\e^{-\Gamma k_1^2(t-t_1)}
\int\limits_0^{t_1}\cd t_2\>\e^{-\Gamma k_2^2(t_1-t_2)}\Delta'[Z-Z_2]
\int\limits_0^{t_2}\cd\tau\>\e^{-\Gamma k_1^2(t_2-\tau)}\Delta''[Z_1-Z(\tau)]
\nonumber\\
&\phantom{=}-
\int_{k_1,k_2}\int\limits_0^t\cd t_1\>\e^{-\Gamma k_1^2(t-t_1)}
\int\limits_0^{t_1}\cd t_2\>\e^{-\Gamma k_2^2(t_1-t_2)}
\Gamma k_1^2\int\limits_0^{t_2}\cd t_3\e^{-\Gamma k_1^2(t_2-t_3)}\Delta'[Z-Z_3]\times
\nonumber\\&\phantom{=}
\int\limits_0^{t_2}\cd\tau\>\e^{-\Gamma k_1^2(t_2-\tau)}\Delta''[Z_1-Z(\tau)]
\end{align}
\end{widetext}
The modification of $D_2^4$ relies on integration by parts for the integral over $t_2$.
Upon inspection of the integrals, one sees, that the integral over $k_1$
in the diagrams $D_1^4$ and $D_2^4$ is well-behaved for $t\to\infty$ in $D>4$,
while the integral over $k_2$ requires only $D>2$. The sum $S=2D_1^4+D_2^4$ on the
other hand behaves vice versa. Of course, in expressions of order $4$ not much is gained
with that, but in higher orders this procedure can be used to understand, why the sum of
all diagrams of a certain order $2q$ involves $q-1$ integrals over momenta that
are finite as $t\to\infty$ in $D>4$ and one that works in $D>2$. Note, that in
equations (\ref{eq:4od1}-\ref{eq:4ospi}) it is important to retain the disorder
correlators, because cancellations have to be exact.

It is now easy to extend the calculations provided in (\ref{eq:4od1}-\ref{eq:4ospi})
to see that $4D^6_1+2D^6_2$ (cf. equations (\ref{eq:6od1}) and (\ref{eq:6od2}))
is a regular expression in $D>4$. The sum $4D^6_1+2D^6_2$ has
the critical dimension for $k_1$, the momentum
associated to the Gau\ss ian pair made of the root and the outermost leaf,
reduced so that it is regular in $D>4$ at the price
that now $k_3$ also needs $D>4$ instead of $D>2$. For reasons of space, as the explicit
expressions for the diagrams are rather huge, we do not provide the full calculation here.

The more general idea about the cancellation among trees goes as follows.
Because too many details have to be ascertained
which tends to result in  a huge load of technicalities, we will merely
sketch the procedure by mentioning all intermediate steps without
proving the implicit claims.
First of all, we note, that any tree $T$ involves at least one distinguished
momentum, we denote it by $k_T$, the integration
over which is regular in $D>2$.
In (\ref{eq:4od1}-\ref{eq:4ospi}) we have seen that it is
possible for any tree $T_q$ of any given order $q$, that
is regular in $D>4$, to find partner trees
$T_q^1,\ldots,T_q^m$ of the same order, all regular in $D>4$, 
such that their sum 
is an expression in which the momentum associated to the root has
the property of $k_T$.
Thus, we will without loss of generality assume that for 
any tree that is regular in $D>4$, $k_T$ is the momentum flowing
out of the root.
All our arguments in the following will also hold for trees for which 
this is not the case, if we repeat it for all partner trees and take the sum.

Let, for some tree $T$ the integration
over some momentum $k_p$ associated to a Gau\ss ian pair $p$ be problematic
in $D>2N_0+2$. 
This can happen if along the line from $p_1$ to $p_2$ there are $N_0$ vertices
which form the root of independent subtrees $\{S_i\}$ or if the momentum flow along
the route from $p_1$ 
to $p_2$ is interrupted by $N_0$ independent (w.r.t. the disorder average)
inner subtrees $\{S_i\}$ and continues at the Gau\ss ian partner of the root of 
those subtrees. The first case corresponds to the diagram $D_1^6$ given by (\ref{eq:6od1})
and the second scenario is exemplified by $D_4^6$ which is depicted in (\ref{eq:6od4})
(both for $N_0=2$).
Of course, also a mixture of both events, $N_0$ in total,
has the same effect, as is illustrated by $D_2^6$ and $D_3^6$.
If, as a kind of induction hypothesis, we assume 
that all independent subtrees $S_1,\ldots,S_{N_0}$ 
are regular in $D>4$, we can describe the scheme how to find all
trees that have to be added to $T$ to yield an expression that is regular
in $D>4$. Let us consider first $N_0=2$. Let $s_1$ be the root of $S_1$
and $s_1'$ be its Gau\ss ian partner vertex in $S_1$. As mentioned above,
the integration over the momentum $k_s$, associated to $(s_1,s_1')$,
is regular in $D>2$.
Without loss of generality, we assume that in $T$, $s_1$ is a vertex on the 
non-interrupted path from
$p_1$ to $p_2$. Then there is another tree $T'$, for which the flow
of $k_p$ along the connection between $p_1$ and $p_2$ is interrupted by the
connection between $s_1$ and $s_1'$. The sum of $T$ and $T'$ is then regular
in $D>4$. 
The integral over $k_p$ in the sum $T+T'$ has decreased its
critical dimension by 2 at the cost, that now the integration
over $k_s$ needs $D>4$ to be bounded for large $t$.

So far, this is the idea, how the cancellation among trees works in $D>4$
to give a regular expression. For a thorough proof, we would have to
give evidence for each single intermediate step.
After all, the practical benefit of such a detailed proof
is little and no further insight can be expected.
Thus, although a rigorous proof has not been established,
a consistent picture of the 
behaviour of the perturbation expansion has emerged. In $D>4$ all perturbative
orders are regular, and in $D\le 4$ our somewhat crude estimate of the
disorder correlator derivatives by constants gives bad results. 

\section{\label{app:hocalc}%
Analysis of the bush graphs%
}
To analyse the term
\begin{align}
T_1&=\int\limits_0^t\cd t_1\cd t_2\>
\Delta[Z_1-Z_2]
\int{\cd^Dk\over (2\pi)^D}\>\e^{-\Gamma k^2(2t-t_1-t_2)}
\end{align}
from equation (\ref{eq:buschallg}),
we start with the decomposition of the function $\Delta[Z(t_1)-Z(t_2)]$
in a double Fourier series in $t_1$ and $t_2$, respectively.
Recall, that $Z(t)=(h/\omega)\sin \omega t$.
\begin{align}
\Delta[Z(t_1)-Z(t_2)]=\int_q\hat\Delta(q)\sum\limits_{m,n}&J_m(qh/\omega)J_n(-qh/\omega)
\times\nonumber\\
&\e^{\ii\omega(mt_1+nt_2)}.
\end{align}
Here, $\hat\Delta$ is the Fourier transform of $\Delta$ and $J_m$ is the Bessel function
of the first kind. For symmetry reasons, only terms with an even value of $m+n$ contribute.
This gives
\begin{align}
\label{eq:t1zerl}
T_1=\int_q\hat\Delta(q)\bigg[&L_{0,0}(q,t)+\sum\limits_{m\ne 0}
L_{2m,0}(q,t)+\nonumber\\
&\sum\limits_{m,n\ne 0}L_{m,n}(q,t)\bigg],
\end{align}
where we have introduced
\begin{align}
\label{eq:lmndef}
L_{m,n}(q,t)=&J_m\left({qh\over\omega}\right)J_n\left(-{qh\over\omega}\right)
\int\limits_0^t\cd t_1\cd t_2\>\e^{\ii\omega(mt_1+nt_2)}\nonumber\\
&\int{\cd^Dk\over(2\pi)^D}\>\e^{-\Gamma k^2(2t-t_2-t_2)}\nonumber\\
=&J_m\left({qh\over\omega}\right)J_{n}\left(-{qh\over\omega}\right)\times\nonumber\\
&\int{\cd^Dk\over(2\pi)^D}\>
{\left[\e^{\ii m\omega t}-\e^{-\Gamma k^2t}\right]\over\Gamma k^2+\ii m\omega}
{\left[\e^{\ii n\omega t}-\e^{-\Gamma k^2t}\right]\over\Gamma k^2+\ii n\omega}.
\end{align}
The behaviour for $t\to\infty$ is dominated by the leading term $L_{0,0}(q,t)$,
which is given by
\begin{align}
L_{0,0}(q,t)&=J_0^2\left({qh\over\omega}\right){S_D\over\Gamma^2(2\pi)^D}\int\limits_0^\Lambda
\cd k\>k^{D-5}\left[1-\e^{-\Gamma k^2t}\right]^2\nonumber\\
&=J_0^2\left({qh\over\omega}\right)\>{t^{4-D\over 2}\over\Gamma^{D\over 2}}\>
A_D(t/\vartheta).
\end{align}
The function $A_D(x)$ has already been introduced in equation (\ref{eq:ADfunc}).
The sub-leading terms are given by
\begin{align}
L_{2m,0}=&J_0\left({qh\over\omega}\right)J_{2m}\left({qh\over\omega}\right)\times\nonumber\\
&\int{\cd^Dk\over(2\pi)^D}\>{\left[1-\e^{-\Gamma k^2t}\right]\over\Gamma k^2}
{\left[\e^{\ii 2m\omega t}-\e^{-\Gamma k^2t}\right]\over\Gamma k^2+\ii 2m\omega}.
\end{align}
Thus,
\begin{align}
\sum\limits_{m\ne 0}&L_{2m,0}(q,t)=\nonumber\\
&\sum\limits_{m=1}^\infty
J_0\left({qh\over\omega}\right)J_{2m}\left({qh\over\omega}\right)
{t^{2-D\over 2}\over\omega\Gamma^{D\over 2}}\alpha_{D}^m(t/\vartheta,\omega t).
\end{align}
Here, the function $\alpha_{D}^m(x,y)$ is given by
\begin{align}
\alpha_D^m(x,y)=&{2S_D\over(2\pi)^D}\int\limits_0^{\sqrt x}
\cd p\>p^{D-3}\left[1-\e^{-p^2}\right]
\times\nonumber\\
&{{p\over y}\cos 2my+2m\sin 2my-{p\over y}\e^{-p}\over (p^2/y^2)+4m^2},
\end{align}
which, for $x\to\infty$ behaves in the same way, as $a_D(x)$ (cf. (\ref{eq:2othetaint})),
independent of $m\ne 0$.

The last term in (\ref{eq:t1zerl}) does not require further consideration. For
$m,n\ne 0$, the function $L_{m,n}$ remains finite as $t\to\infty$. There is no
infrared problem with the $k$-integral in equation (\ref{eq:lmndef}) any more.
Recalling the two time scales $\tau$ and $\vartheta$, that we have encountered
before (cf. (\ref{eq:tautscale}) and (\ref{eq:thetatscale}))
we thus have
\begin{align}
{u^2\over\ell^2}T_1=&
\left({t\over\tau}\right)^{4-D\over 2}
\kappa_D\left({t\over\vartheta}\right)+
\nonumber\\
&\left({t\over\tau}\right)^{4-D\over 4}
{u\Lambda^{D\over 2}\over\omega\ell}
k_D\left({t\over\vartheta},\omega t\right)+
\nonumber\\
&{u^2\Lambda^D\over\omega^2\ell^2}
P_D\left({t\over\vartheta},\omega t\right).
\end{align}
Hereby, we have introduced
\begin{align}
\label{eq:kappad}
\kappa_D\left({t\over\vartheta}\right)=&
A_D(t/\vartheta)\int_q\hat\Delta(q)J_0^2\left({qh\over\omega}\right)\\
k_D\left({t\over\vartheta},\omega t\right)=&
\left({\vartheta\over t}\right)^{D\over 4}\sum\limits_{m=1}^\infty
\alpha_D^m\left({t\over\vartheta},\omega t\right)\times\nonumber\\
&\int_q\hat\Delta(q)
J_0\left({qh\over\omega}\right)J_{2m}\left({qh\over\omega}\right)\\
P_D\left({t\over\vartheta},\omega t\right)=&
{\omega^2\over\Lambda^D}
\sum\limits_{m,n\ne 0}\int_q\hat\Delta(q)L_{m,n}(q,t).
\end{align}

The second factor in (\ref{eq:buschallg})
\begin{align}
T_2=\int\limits_0^t\cd t'\>\Delta^{(2p-1)}[Z-Z']
\int{\cd^Dk\over (2\pi)^D}\>\e^{-\Gamma k^2(t-t')}
\end{align} 
can be treated in the same way, like the
second order graph in section \ref{sec:walls:perturbation}, 
by splitting off the Fourier-0-mode
\begin{align}
\Delta^{(2p-1)}[Z(t)-Z(t')]&={F_0^{[2p-1]}(\omega t)+p(t,t')\over\ell^{2p-1}}.
\end{align}
Following the calculations in section \ref{sec:walls:perturbation}, with
$F_0(\omega t)$ replaced by $F_0^{[2p-1]}(\omega t)$, we arrive at
\begin{align}
{u^2\ell^{2(p-1)}\over\omega\ell}T_2=&
{u\Lambda^{D\over 2}\over\omega\ell}
\left({t\over\tau}\right)^{4-D\over 4}
f_D\left({t\over\vartheta},\omega t\right)+
\nonumber\\
&{u^2\Lambda^D\over\omega^2\ell^2}
p_D\left({t\over\vartheta},\omega t\right),
\end{align}
where
\begin{align}
f_D\left({t\over\vartheta},\omega t\right)=&
\left({\vartheta\over t}\right)^{D\over 4}
a_D(t/\vartheta)F_0^{[2p-1]}(\omega t)\\
p_D\left({t\over\vartheta},\omega t\right)=&
{S_D\over(2\pi)^D}\int\limits_0^{\sqrt{t/\vartheta}}\cd p\>p^{D-1}\e^{-p^2}
\int\limits_0^t\cd t'\>{p(t,t-t')\over[\Gamma t']^D}.
\end{align}
The integral over $p$ in $p_D$ is certainly convergent for any $D$ in
the limit $t\to\infty$, and the
integral over $t'$ converges for any $D>0$, since $p(t,t-t')$ is a bounded 
oscillation around zero (without zero Fourier mode) in $t'$.

\section{\label{app:width}%
The width of ac-driven interfaces%
}
Apart from the velocity of the mean position of an interface
in a random potential, there is another interesting quantity that 
deserves investigation: the mean square deviation of a given realisation
from the mean. More precisely, we refer to the quantity
\begin{equation}
w=\disav{(\disav{z}-z)^2}.
\end{equation}
In the first order of the perturbation expansion, $w$ reads
\begin{align}
w=\disav{(\disav{Z+u\zeta_1}-Z-u\zeta_1)^2}=u^2\disav{\zeta_1^2}+{\cal O}(u^4).
\end{align}
In the case of infinitely extended interfaces, this quantity measures thus
the typical width of the interface.

So, for infinitely extended domain walls, the typical width to first order in
perturbation theory is given by (cf. (\ref{eq:w1storder}))
\begin{align}
\label{eq:wwidth}
\disav{\zeta_1^2}(x,t)=\int\limits_0^t\cd t_1\cd t_2\>&\Delta[Z(t_1)-
Z(t_2)]\times\nonumber\\
&\int{\cd^D k\over(2\pi)^D}\>\e^{-\Gamma k^2(2t-t_1-t_2)}.
\end{align}
Comparing this to $T_1$, given by equation (\ref{eq:t1t2def}), we find
$\disav{\zeta_1^2}=T_1$.
Using equation (\ref{eq:t1absch}), we thus have for the
asymptotics $t\to\infty$
\begin{align}
u^2\disav{\zeta_1^2}\sim\ell^2\left[{t\over\tau}\right]^{4-D\over 2}A_D(t/\vartheta)
\end{align}
The function $A_D$, given by (\ref{eq:ADfunc}), remains bounded for $D<4$ and
grows logarithmically in its argument in case $D=4$. 
Thus, the growth of the perturbative estimate of $w$ in time is given by the prefactor
$t^{4-D\over 2}$ and $\log t$ for $D<4$ and $D=4$, respectively.

So, in contrast to the first order perturbative result for the interface's
velocity, which remains finite as $t\to\infty$ for $D>2$, the width of the
interface indicates the correct critical dimension $D=4$ already to first order.

\section{\label{app:reg}%
Regularity of the mean-field perturbation expansion%
}
In section \ref{sec:mf:pert:consistency} we have analysed, how the unbounded contributions,
contained in the two diagrams that involve a curly line, mutually cancel in the
second non-vanishing perturbative order. In this appendix, 
we are going to explain how this cancellation process generalises to all orders in 
perturbation theory. As before, for simplicity, 
we work with the diagrams for the disorder-averaged
velocity, that arise by
just removing the curly lines from the root of the diagrams for $\diszeta$
(cf. equation (\ref{eq:diagexpand})). 
In a velocity diagram contributing to the $n$-th order (recall, that only for even $n$
the corrections are non-zero), 
any curly line connects two trees of order $p$ and $q$ (both even) 
with the restriction $p+q=n$.
Both trees appear in the expansion of lower orders, 
namely $p$ and $q$, respectively.
In the following, we sketch an inductive proof for the claim
that the unbounded terms
originating from trees with curly internal lines cancel among each other.

Let us assume, that for order $n$ we have achieved to ensure regularity. 
For every unbounded tree $T$, there is thus a set $T^1,\ldots,T^a$ of, let us call them
\emph{cancelling trees}, such that $T+T^1+\ldots+T^a$ is a regular, 
bounded expression in time.
As a starting point for the induction, take $n=4$, where the validity of
the claim has been verified in section \ref{sec:mf:pert:consistency}.
It is now the task to validate the regularity for order $n+2$. 
First of all, we consider the process of attaching the root of a
regular tree $S$ (with no internal curly line) of order $s$ 
by a curly line to a vertex $v$ of another regular tree $R$ of order $r=n+2-s$ 
to obtain a new irregular tree $A$ of order $n+2$.
The vertex $v$ must be connected to another vertex $w\in R$ by a dashed line, to
carry out the Gau\ss ian disorder average. Without loss of generality, we assume
that $v$ is connected to $w$ by a path that first makes a step towards the root.  
The rules for the diagrammatic
expansion ensure, that there is a maximal regular subtree $T\subset R$, 
which contains $v$ and $w$.

Using partial integration, it is possible to move the vertex to which $S$ is connected
(via the curly line) to a neighbouring vertex in $T$. Thus, it is possible to
move the connection vertex along the unique way (in $T$) from $v$ to $w$. 
We are going to show, that once $w$ is reached, we have
obtained the cancelling tree which is unique. 
Diagrammatcially, the
process of moving the connection vertex from $v$ to $w$ reads:
\begin{eqnarray}
\parbox{21mm}{%
\includegraphics{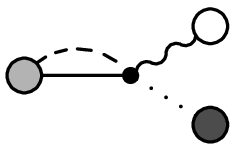}%
}
&=&
\>\>\parbox{21mm}{%
\includegraphics{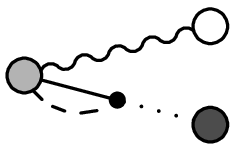}%
}+D
\end{eqnarray}
Here, the blank circle represents $S$, the lightgrey circle stands for the subtree $R_1$ of 
$R$, to which $v$ connects and the 
darkgrey shaded circle denotes trees which run out of $v$ 
(summarised in the following as $R_2$). Certainly, in general there 
may be dashed lines between the dark- and the lightgrey circle, which we have omitted as
they are not relevant for the forthcoming discussion.
The dotted line just serves as a joker - it is not important to specify how many lines
go out of $v$. 
The last term $D$ collects the left-over terms
from the partial integration. Note, that, if it takes several steps to go from $v$ to $w$,
the intermediate expressions (in the partial integration) are not in accordance with
our diagrammatic rules, because the order of derivative of the disorder correlators
does not appear correctly (it remains the same but the graph has changed). 
Keeping this small peculiarity in mind, it is nevertheless
instructive to think in diagrammatic terms.

To illustrate the procedure, we take a look at the first step:
\begin{widetext}
\begin{align}
\parbox{21mm}{%
\includegraphics{cdiag1.eps}%
}
=&R_1(t)\int\limits_0^{T_1}\cd t_1\>\e^{-c(T_1-t_1)}(-1)^\nu\Delta^{(\mu+\nu)}[Z(\tau)-Z(t_1)]
R_2(t_1)\int\limits_0^{t_1}\cd t_2 S(t_2)\\
=&R_1(t)\int\limits_0^{T_1}\cd t_2 S(t_2)
\int\limits_0^{T_1}\cd t_1\e^{-c(T_1-t_1)}(-1)^\nu\Delta^{(\mu+\nu)}[Z(\tau)-Z(t_1)]R_2(t_1)
\nonumber\\
&-R_1(t)\int\limits_0^{T_1}\cd t_1\>\e^{-c(T_1-t_1)}S(t_1)\int\limits_0^{t_1}
\cd t_2\>\e^{-c(t_1-t_2)}(-1)^\nu\Delta^{(\mu+\nu)}[Z(\tau)-Z(t_2)]R_2(t_2)
\end{align}
\end{widetext}
The order of the derivative (i.e. the number of outgoing lines) of $w$ and $v$ are denoted 
by $\mu$ and $\nu$, respectively.
The time, at which the whole diagram is to be evaluated, is $t$, the time corresponding to
the vertex to which $v$ is connected is given by $T_1$, $t_1$ is thus the time associated 
to $v$ and so on. The time of $w$ is $\tau$. 
Thus, we see, that if $w$ is not the vertex to which $v$ is directly connected 
(then $T_1\ne\tau$ in general),
the first expression after partial integration cannot be a valid diagram: $v$ has lost
one order of derivative ($\nu-1$ lines go out instead of $\nu$), but the derivative
of the correlator $\Delta$ has not changed. A valid diagram however reappears, when the
connection of $S$ has reached $w$. Then, $v$ has lost an outgoing line, but $w$ received
one more and we indeed have achieved a cancelling tree:
the factor $(-1)^\nu$ remains, the true diagram, however, has $(-1)^{\nu-1}$. The signs are
different, thus the two trees cancel. 
The left-over term from the partial integration is again regular. This can be seen because
all time integrals carry an exponential damping term. It is clear, that this is generally 
true for every partial integration step.

To go one step further, we assume now $S$ to be irregular.
Essentially, the same procedure works, but there are more
cancelling trees: one has take all cancelling trees $\{S^i\}$ for $S$ into account 
(which exist by induction hypothesis), thus
$S$ is replaced by $\sum S^i$ and thence the left-over terms are again regular.

A possible irregularity of $R$ can be accounted for in the same way. It is, however,
important to explain why this is possible, i.e. what are $v$ and $w$ in the cancelling
trees for $R$. In the case of irregular $S$ the problem was easy, since all
trees have a unique root. As we have seen already, the procedure of creating cancelling
trees does not change the structure of regular subtrees. Thence, all cancelling trees for 
$R$ contain $T$. This makes clear, which $v$ and $w$ have to be chosen in the cancelling
trees: they are well-defined in $T$ and $T$ is a well-defined subtree of the cancelling
trees. Thus, repeating the whole procedure described above for all cancelling trees of $R$
yields the complete set of cancelling trees for $A$ in the most general setting.

\section{\label{app:symbols}%
List of recurrent symbols%
}
\begin{table*}
\begin{ruledtabular}
\begin{tabular}{ccc}
Symbol&Quantity&Reference\\
\hline
$z(x,t)$     & interface profile                    & Sec.\ref{sec:walls:model}\\
$D$          & internal interface dimension         & Sec.\ref{sec:walls:model}\\
$h$          & amplitude of the driving force       & Eq.(\ref{eq:acdrive}) and
                                                      Eq.(\ref{eq:weom})\\
$\omega$     & frequency of the ac-driving force    & Eq.(\ref{eq:acdrive})\\
$\Gamma$     & elastic stiffness of the interface   & Eq.(\ref{eq:weom})\\
$c$          & elasticity constant (mean-field)     & Sec.\ref{sec:mf:model} and
                                                      Eq.(\ref{eq:mfgeom})\\
$\ell$       & disorder correlation length          & Sec.\ref{sec:walls:model} and 
                                                      Eq.(\ref{eq:discorr})\\
$u$          & disorder strength                    & Eq.(\ref{eq:weom})\\
$\eta$       & disorder strength (mean-field)       & Sec.\ref{sec:mf:model} and
                                                      Eq.(\ref{eq:weom})\\
$\Delta$     & disorder correlator                  & Sec.\ref{sec:walls:model} and 
                                                      Eq.(\ref{eq:discorr}) and
                                                      Sec.\ref{sec:mf:model}\\
$g$          & disorder configuration               & Eq.(\ref{eq:weom}) and
                                                      Eq.(\ref{eq:mfgeom})\\
$\larkin$    & Larkin length                        & Eq.(\ref{eq:larkin})\\
$Z(t)$       & solution in absence of disorder      & Sec.\ref{sec:walls:perturbation} and
                                                      Sec.\ref{sec:mf:pert}\\
$\zeta$      & disorder correction to the solution  & Sec.\ref{sec:walls:perturbation} and
                                                      Sec.\ref{sec:mf:pert}\\
$G$          & propagator                           & Eq.(\ref{eq:pertprop}) and
                                                      Eq.(\ref{eq:mfprop})\\
$\Lambda$    & inverse smallest length scale
               (UV-cutoff)                          & Sec.\ref{sec:walls:perturbation}\\
$\tau$       & roughening time                      & Sec.\ref{sec:walls:dl4} and
                                                      Eq.(\ref{eq:tautscale})\\
$\vartheta$  & additional transience time scale     & Sec.\ref{sec:walls:dl4} and
                                                      Eq.(\ref{eq:thetatscale})
\end{tabular}
\end{ruledtabular}
\end{table*}

\bibliography{wand}

\end{document}